\documentclass[aps,prb,twocolumn,superscriptaddress,floatfix]{revtex4}

\usepackage{graphicx}
\usepackage{dcolumn}
\usepackage{amsmath}
\usepackage{amsfonts}
\usepackage{amssymb}
\usepackage{bm}
\usepackage{color}

\begin{document}

\newcommand{\ds}{\displaystyle}
\newcommand{\sigmaop}{\hat{\sigma}}
\newcommand{\aop}{\hat{a}}
\newcommand{\adop}{\hat{a}^\dagger}
\newcommand{\Hop}{\hat{H}}

\preprint{F. Deppe {\it et al.}}
\title{Phase-Coherent Dynamics of a Superconducting Flux Qubit with Capacitive-Bias Readout}

\author{F.~Deppe}
 \affiliation{Walther-Mei{\ss}ner-Institut,
              Walther-Mei{\ss}ner-Str.~8, D-85748 Garching, Germany}
 \affiliation{NTT Basic Research Laboratories, NTT Corporation, Kanagawa,
              243-0198, Japan}
 \affiliation{CREST, Japan Science and Technology Agency, Saitama, 332-0012
              Japan}

\author{M.~Mariantoni}
 \affiliation{Walther-Mei{\ss}ner-Institut,
              Walther-Mei{\ss}ner-Str.~8, D-85748 Garching, Germany}

\author{E.~P.~Menzel}
 \affiliation{Walther-Mei{\ss}ner-Institut,
              Walther-Mei{\ss}ner-Str.~8, D-85748 Garching, Germany}

\author{S.~Saito}
 \affiliation{NTT Basic Research Laboratories, NTT Corporation, Kanagawa,
              243-0198, Japan}
 \affiliation{CREST, Japan Science and Technology Agency, Saitama, 332-0012
              Japan}

\author{K.~Kakuyanagi}
 \affiliation{NTT Basic Research Laboratories, NTT Corporation, Kanagawa,
              243-0198, Japan}
 \affiliation{CREST, Japan Science and Technology Agency, Saitama, 332-0012
              Japan}

\author{H.~Tanaka}
 \affiliation{NTT Basic Research Laboratories, NTT Corporation, Kanagawa,
              243-0198, Japan}
 \affiliation{CREST, Japan Science and Technology Agency, Saitama, 332-0012
              Japan}

\author{T.~Meno}
 \affiliation{NTT Advanced Technology, NTT Corporation, Kanagawa, 243-0198,
              Japan}

\author{K.~Semba}
 \affiliation{NTT Basic Research Laboratories, NTT Corporation, Kanagawa,
              243-0198, Japan}
 \affiliation{CREST, Japan Science and Technology Agency, Saitama, 332-0012
              Japan}

\author{H.~Takayanagi}
 \affiliation{CREST, Japan Science and Technology Agency, Saitama, 332-0012
              Japan}
 \affiliation{Tokyo University of Science, 1-3 Kagurazaka, Shinjuku,
              Tokyo 162-8601, Japan}

\author{R.~Gross}
 \affiliation{Walther-Mei{\ss}ner-Institut,
              Walther-Mei{\ss}ner-Str.~8, D-85748 Garching, Germany}

\begin{abstract}
We present a systematic study of the phase-coherent dynamics of a superconducting three-Josephson-junction flux qubit. The qubit state is detected with the integrated-pulse method, which is a variant of the pulsed switching DC~SQUID method. In this scheme the DC~SQUID bias current pulse is applied via a capacitor instead of a resistor, giving rise to a narrow band-pass instead of a pure low-pass filter configuration of the electromagnetic environment. Measuring one and the same qubit with both setups allows a direct comparison. With the capacitive method about four times faster switching pulses and an increased visibility are achieved. Furthermore, the deliberate engineering of the electromagnetic environment, which minimizes the noise due to the bias circuit, is facilitated. Right at the degeneracy point the qubit coherence is limited by energy relaxation. We find two main noise contributions. White noise is limiting the energy relaxation and contributing to the dephasing far from the degeneracy point. $1/f$-noise is the dominant source of dephasing in the direct vicinity of the optimal point. The influence of $1/f$-noise is also supported by non-random beatings in the Ramsey and spin echo decay traces. Numeric simulations of a coupled qubit-oscillator system indicate that these beatings are due to the resonant interaction of the qubit with at least one point-like fluctuator, coupled especially strongly to the qubit.
\end{abstract}

\date{\today}


\maketitle

\section{Introduction}
\label{sec:Introduction}

In the past two decades it has become evident that quantum mechanics can give rise to new fascinating possibilities for communication and information processing. Based on quantum two-level systems (qubits) instead of conventional classical bits, secure quantum communication protocols and quantum algorithms have been proposed. The latter promise a significant speed-up of certain computational tasks, such as the factorization of large numbers, by making use of a massive quantum parallelism and entanglement\cite{Deutsch:1985a,Deutsch:1992a,Shor:1994a,Grover:1996a,Simon:1997a,Shor:1997a,Nielsen:2000a}. Subsequently, there have been intensive research efforts aiming at the development of adequate hardware concepts\cite{DiVincenzo:2000a}. The successful implementation of qubits has first been proposed and demonstrated for microscopic systems (e.g., NMR\cite{Vandersypen:2001a}, trapped ions\cite{Cirac:1995a,Leibfried:2003a},  and cavity-QED systems\cite{Raimond:2001a,Monroe:2002a,Mabuchi:2002a,Cirac:2004a,Walther:2006a}). Although these systems possess sufficiently long decoherence times, they have drawbacks regarding their scalability to large architectures, which are required for practical quantum computing. In contrast to that, scalability is a specific advantage of solid-state systems. Therefore, solid-state quantum circuits have attracted increasing interest giving rise to a rich field of theoretical and experimental investigations in the context of quantum information processing. The fabrication process of solid-state quantum circuits is strongly facilitated by the use of well established techniques from micro- and nanoelectronics such as lithography and thin-film technology. Obviously, this results in a high degree of tunability and potential scalability to larger units. Furthermore, quantum systems clearly have a great potential in solid-state electronics because the ongoing miniaturization of integrated circuits requires the deliberate use of quantum-mechanical effects in the near future.

Among solid-state systems, superconducting devices are particularly interesting. Quantum information processing based on superconducting circuits exploits the intrinsic coherence of the superconducting state. In fact, this state is separated by an energy gap from the normal conducting states and, therefore, relatively long coherence times can be achieved. In general, superconducting quantum circuits consist of a single or multiply-connected superconducting lines intersected by Josephson junctions. Quantum information can be stored in the number of superconducting Cooper pairs (e.g., the charge qubit\cite{Nakamura:1999a,Nakamura:2001a,Yamamoto:2003a} and the quantronium\cite{Vion:2002a,Ithier:2005a}), in the direction of a circulating persistent current (e.g., the three-junction flux qubit\cite{Mooij:1999a,Orlando:1999a,VanderWal:2000a,Chiorescu:2003a,Yoshihara:2006a,Kakuyanagi:2007a} and the RF-SQUID-based flux qubit\cite{Friedman:2000a}) or in oscillatory states (e.g., the phase qubit\cite{Martinis:2002a,Yu:2002a,Martinis:2003a,Simmonds:2004a,Cooper:2004a,Martinis:2005a,Steffen:2006a}). To a large extent, superconducting qubits are protected from the unwanted interaction with environmental degrees of freedom by the superconducting gap. Furthermore, design-dependent internal symmetries can reduce the influence of noise arising from the control and readout circuitry. Nevertheless, the uncontrolled loss of coherence still represents a key issue. In the most simple theoretical description\cite{Bloch:1953a,Redfield:1957a} decoherence in a quantum two-level system is described in terms of two rates or times. The first is the longitudinal or energy relaxation rate $\Gamma_1\equiv T_1^{-1}$, which describes the transitions between the two qubit states due to high-frequency fluctuations. The second is the transverse or dephasing rate $\Gamma_\varphi\equiv T_\varphi^{-1}$, which describes the loss of phase coherence within the same qubit state caused by low-frequency noise. While $\Gamma_1$ can be measured directly, $\Gamma_\varphi$ has to be extracted from the experimentally accessible quantity $\Gamma_2=\Gamma_1/2+\Gamma_\varphi$.

The coherence time of a quantum state is one of the key figures of merit for practical quantum information devices. In fact, it determines the number of qubit operations that can be performed without errors. The presently achieved coherence times in superconducting qubits (ranging from approximately $10\,$ns to $5\,\mu$s) are not sufficient for the realization of more complex quantum circuits. For this reason, decoherence due to the coupling of the qubit to the environmental degrees of freedom or due to noise introduced by the control and readout circuitry has been addressed in several recent experimental\cite{Nakamura:2002a,Astafiev:2004a,Schuster:2005a,Martinis:2005a,Bertet:2005a,Ithier:2005a,Astafiev:2006a,Yoshihara:2006a,Kakuyanagi:2007a} and theoretical studies\cite{Makhlin:2001a,Shnirman:2002a,Makhlin:2004a,Ithier:2005a,Shnirman:2005a}. However, a thorough understanding of the decoherence processes and detailed knowledge of the origin of the noise sources is still missing, making the further clarification of possible sources of decoherence highly desirable. In this sense, a promising path to follow is to study the dependence of the characteristic decay times (energy relaxation and dephasing) on external parameters such as the applied magnetic flux in superconducting flux qubits\cite{Bertet:2005a,Yoshihara:2006a,Kakuyanagi:2007a}. 

It is noteworthy to mention that the interaction of superconducting qubits with microwave resonators has recently attracted increasing attention. It has turned out that the qubit-resonator interaction is the circuit equivalent of the atom-photon interaction in optical cavity QED\cite{Raimond:2001a,Monroe:2002a,Mabuchi:2002a,Cirac:2004a,Walther:2006a}. In other words, the formalism of cavity QED has been successfully transferred to the realm of superconducting systems, which can now be treated with the standard tools developed in quantum optics. This new field is referred to as circuit QED and has given rise to a set of important method for the manipulation and dispersive readout of qubits\cite{Blais:2004a,Chiorescu:2004a,Wallraff:2004a,Wallraff:2005a,Johansson:2006a,Mariantoni:2006a}. Particularly interesting for the purposes of the present work are the investigations on coherent dynamics presented in Refs.~\onlinecite{Morigi:2002a} and \onlinecite{Meunier:2005a}. Together with our work they allow for further insight in the issue of nearly-resonant spurious resonators or two-level systems interacting with superconducting qubits\cite{Simmonds:2004a}.

In this paper, we present a detailed study of both energy relaxation and dephasing rates of a superconducting three-Josephson-junction flux qubit as a function of the applied magnetic flux close to the degeneracy point. For this type of qubit several different readout methods have been proposed and successfully implemented. They range from the simple switching-DC-SQUID method\cite{Mooij:1999a} to more sophisticated techniques such as the inductive readout\cite{Zorin:2002a,Greenberg:2002a,Lupascu:2004a,Lupascu:2005a} or the bifurcation amplifier\cite{Siddiqi:2004a,Siddiqi:2005a,Siddiqi:2006a,Lupascu:2006a}. The two latter have the potential to achieve quantum non-demolition measurements\cite{Lupascu:2007a} and very large signal visibility, which are very important features for future applications. In this study we have chosen a switching DC~SQUID readout, which enables us to study the effect of two fundamentally different electromagnetic environments (capacitive and resistive) on the decoherence of one and the same flux qubit easily. Furthermore, the switching-DC-SQUID readout is attractive because of its simple technical implementation (e.g., no cold amplifiers or other sophisticated high-frequency circuits are needed). In our measurements of the coherent dynamics of a superconducting flux qubit we apply the bias pulse to the DC~SQUID detector via a capacitor ($C$-Bias). Such a setup reduces low-frequency bias current noise. We experimentally demonstrate that the decoherence of our qubit at the optimal point is dominated by relaxation, i.e., by high-frequency noise. Nevertheless, we still find the signature of $1/f$-noise in the vicinity of the optimal point. We compare our results to those obtained from resistive-bias ($R$-Bias) measurements on the same qubit and find a similar $1/f$ spectral noise density, despite the different bias line filtering conditions. This demonstrates that low-frequency noise from the bias line is not the dominating noise source in our system. The relaxation time at the degeneracy point is slightly reduced for the capacitive method. We are able to exclude the different bias line configuration as the reason for this $T_1$-time reduction. Instead, we attribute it to a change in the high-frequency setup when switching between the two methods. Furthermore, we show that the resonant interaction of the investigated flux qubit with two-level fluctuators is responsible for non-random beatings in the measured Ramsey and spin echo traces. Together with the observed weak influence of low-frequency bias noise on the qubit coherence, this suggests that an ensemble of fluctuators on the sample chip and close to the qubit is the main source of the measured $1/f$-noise. Additionally, we find that in our setup, compared to the resistive-bias measurements, the capacitive-bias method allows for four times faster switching pulses and helps to obtain an increased visibility.

The paper is organized as follows. In Sec.~\ref{sec:Thesuperconductingquantumcircuit}, we first introduce the fabrication process and the basic properties of the investigated three-junction flux qubit, the experimental setup, and the general measurement concepts. Then, in Sec.~\ref{sec:Capacitivebias:Theintegratedpulsemethod}, we describe the details of the capacitive-bias readout scheme, which is based on the switching of a DC~SQUID. Here, in particular the differences to the usual resistive-bias setup are discussed. In Sec.~\ref{sec:Experimentalresultsanddiscussion}, we show the results of the capacitive-bias measurements on the phase coherent dynamics of our flux qubit. We present a detailed quantitative analysis of the results of these measurements, which have been performed both in the frequency (microwave spectroscopy) and in the time domain (energy relaxation, Ramsey fringes, and spin echo decay).  We discuss the effect of the different electromagnetic environments by way of comparing the capacitive-bias results to those obtained from resistive-bias measurements on the same qubit. In this way, we avoid possible ambiguities arising from a comparison of different samples due to the spread in fabrication and/or the rich variety of different decoherence sources. From the measured flux dependence of the energy relaxation and dephasing rates we are able to quantify the contributions of $1/f$- and white noise to the total flux noise spectral density affecting the qubit. Finally, in Sec.~\ref{sec:SpinechoandRamseybeatings}, we discuss the origin of non-random beatings in the Ramsey and spin echo decay traces. We show that these beatings cannot be caused by imperfect control pulses or undesired probe frequency detuning because in our experiments we apply the so-called phase-cycling method\cite{Kakuyanagi:2007a}. Comparison with numerical simulations strongly suggests that these beatings originate from the interaction of the flux qubit with one (or a few) point-like harmonic oscillators or two-level fluctuators.

\section{The superconducting quantum circuit}
\label{sec:Thesuperconductingquantumcircuit}

In this section, we introduce the working principle and characteristic parameters of our work-horse, the superconducting three-Josephson-junction flux qubit. We introduce the pulsed switching DC~SQUID readout technique, placing special emphasis on the capacitive-bias method. Furthermore, the control of the qubit state with microwave pulses as well as the filtering of the measurement lines are discussed.

\subsection{The superconducting three-Josephson-junction flux qubit}
\label{subsec:Thethree-Josephson-junctionfluxqubit}

\begin{figure}[b]
\centering{\includegraphics[width=86mm]{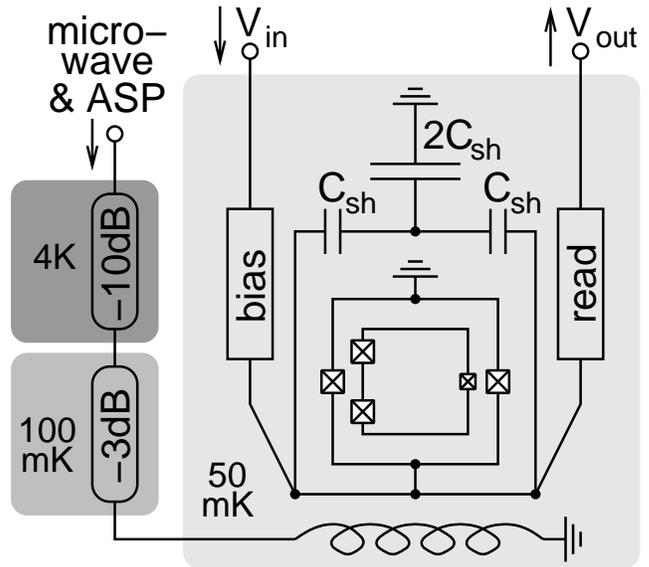}}

\caption{Sketch of the sample layout. The Al/AlO$_x$/Al Josephson junctions
         of qubit and DC~SQUID are represented by the symbol $\boxtimes$.
         $V_{\rm in}$ and $V_{\rm out}$ are the bias pulse and the switching
         signal of the DC~SQUID, respectively . $C_{\rm sh}$ is the effective
         shunting capacitance  of the DC~SQUID.  The boxes titled ``bias''
         and ``read'' represent the circuit elements used to engineer
         the electromagnetic environment of the qubit. They are located
         partially on-chip and partially on the PCB surrounding the
         sample chip. ``microwave~\&~ASP'' denote the microwave control
         pulse pattern and the adiabatic shift pulse, which are attenuated
         at low temperature and coupled to the qubit via an on-chip antenna 
         (coiled shape).}
 \label{Deppe_PRB2007_Fig1}
\end{figure}

The investigated sample is fabricated on a $3.5\times3.5\,\text{mm}^2$ SiO$_2$-passivated Si substrate. Standard electron beam lithography and aluminum thin-film technology are used. The sample chip is surrounded by a T-shaped printed circuit board (PCB). Both sample chip and PCB are placed inside a gold-plated copper box, which is connected to the mixing chamber stage of a dilution refrigerator. All experiments are performed at the base temperature of the fridge, which is approximately $50\,$mK. In the following, we briefly describe the main components of the sample and summarize their characteristic parameters and function. A sketch of the sample layout is shown in Fig.~\ref{Deppe_PRB2007_Fig1}.

The three-Josephson-junction flux qubit\cite{Mooij:1999a} consists of a square-shaped superconducting aluminum loop interrupted by three nano scale Al/AlO$_x$/Al Josephson junctions. The area of two of these junctions is chosen to be the same ($0.03\,\mu\textrm{m}^2$), whereas the third one is designed to be smaller by a factor $\alpha_{\rm design}=0.7$. The qubit (and DC~SQUID) junctions are fabricated using the shadow evaporation technique. The Josephson junctions are characterized by their Josephson coupling energy $E_{\rm J}=I_{\rm c}\Phi_0/2\pi$ and their charging energy $E_{\rm C}=e^2/2C_{\rm J}$. Here, $\Phi_0$ is the flux quantum, $e$ the electron charge, and $I_{\rm c}$ and $C_{\rm J}$ are the critical current and capacitance of the junction, respectively. The specific capacitance of our junctions was determined to be\cite{Deppe:2004a} $C_{\rm s}=100\pm25\,\text{fF}/\mu\text{m}^2$. When the ratio $E_{\rm J}/E_{\rm C}$ is approximately $50$, the device is expected to behave as an effective quantum two-level system (qubit).

The effective Hamiltonian of a flux qubit can be written as \cite{Mooij:1999a,Makhlin:2001a,CohenTannoudji:1977a}
\begin{equation}
   \Hop_{\rm qb}=\frac{\epsilon(\Phi_{\rm x})}{2}
                 \sigmaop_z+\frac{\Delta}{2}\sigmaop_x\;,
   \label{eqn:qubithamiltonian}
\end{equation}
where $\sigmaop_z$ and $\sigmaop_x$ are the Pauli matrices. For vanishing tunnel coupling ($\Delta=0$) the two qubit states correspond to the classical states $\left|{\rm R}\right>$ and $\left|{\rm L}\right>$ with clockwise and counterclockwise persistent currents $I_{\rm p}=\pm I_{\rm c} \sqrt{1-(2\alpha)^{-2}}$ circulating in the loop\cite{Mooij:1999a}. They are separated by the flux-dependent energy $\epsilon(\Phi_{\rm x})=2I_{\rm p}\delta\Phi_{\rm x}$, where $\delta\Phi_{\rm x}\equiv\Phi_{\rm x}-\Phi_{(n)}$ is the applied magnetic flux bias relative to the degeneracy point $\Phi_{(n)}\equiv\left(n+\frac{1}{2}\right)\Phi_0$ and $n$ is an integer. Our sample is designed to be operated close to $\Phi_{\rm x}\simeq\Phi_{(1)}$. For finite coupling ($\Delta>0$), we obtain superpositions of $\left|{\rm R}\right>$ and $\left|{\rm L}\right>$. This results in new qubit eigenstates $\left|0\right>$ and $\left|1\right>$ whose energy difference $E_{01}=\sqrt{\epsilon(\Phi_{\rm x})^2+\Delta^2}=\sqrt{(2I_{\rm p}\delta\Phi_{\rm x})^2+\Delta^2}$ has a hyperbolic flux dependence. At the degeneracy point ($\delta\Phi_{\rm x}=0$, $\epsilon(\Phi_{\rm x})=0$) the qubit is protected from dephasing because $E_{01}$ is stationary with respect to small variations of the control parameter $\delta\Phi_{\rm x}$. Therefore, this point represents the optimal point for the coherent manipulation of the qubit. Also, the qubit eigenstate at the degeneracy point is an equal superposition of $\left|{\rm L}\right>$ and $\left|{\rm R}\right>$, i.e., the expectation value of the persistent current vanishes. Far away from the degeneracy point ($\epsilon(\Phi_{\rm x}) \gg \Delta$) the qubit behaves as a classical two-level system. This example clearly shows the flexibility offered by solid-state-based qubits due to their high degree of tunability. As detailed in Sec.~\ref{sec:Experimentalresultsanddiscussion}, for our device we find $\Delta/h\simeq4\,$GHz, $E_{\rm J}/E_{\rm C} \simeq 50$, and a critical current density $J_{\rm c}\simeq1300\,\text{A/cm}^2$. This means that at an operation temperature $T\simeq50\,$mK the condition $k_{\rm B}T\ll\Delta$ required for the observation of quantum effects is well satisfied.

When describing the influence of fluctuations $\delta\boldsymbol{\omega}$ on the qubit (cf.\ Sec.~\ref{sec:Experimentalresultsanddiscussion}), it is convenient to express the Hamiltonian of Eq.~(\ref{eqn:qubithamiltonian}) in a two-dimensional Bloch vector representation: $\Hop_{\rm qb}=\hbar\boldsymbol{\omega\sigmaop}/2$. Here, $\boldsymbol{\sigmaop}\equiv(\sigmaop_\perp,\sigmaop_\parallel)=(\sigmaop_x,\sigmaop_z)$ and $\boldsymbol{\omega}\equiv(\omega_\perp,\omega_\parallel)=\hbar^{-1}\big(\Delta,\epsilon(\Phi_{\rm x})\big)$. The representation of the Bloch vector $\boldsymbol{\omega}$ in the qubit energy eigenbasis is obtained by multiplying $\boldsymbol{\omega}$ with the rotation matrix $D\equiv\left(\begin{matrix}\cos\theta&-\sin\theta\\\sin\theta&\cos\theta\end{matrix}\right)$ from the left. The Bloch angle $\theta$ is defined\cite{CohenTannoudji:1977a} via $\tan\theta\equiv\Delta/\epsilon(\Phi_{\rm x})$. This results in $\sin\theta=\Delta/\nu_{01}$ and $\cos\theta=\epsilon(\Phi_{\rm x})/\nu_{01}$, where $\nu_{01}\equiv\omega_{01}/2\pi\equiv E_{01}/h=\sqrt{\Delta^2+\epsilon(\Phi_{\rm x})^2}/h$ is the qubit transition frequency. At the degeneracy point the angular transition frequency becomes $\omega_\Delta\equiv\Delta/\hbar$.

\subsection{Manipulation of the qubit state}
\label{subsec:Manipulationofthequbitstate}

The qubit control is achieved by varying the control parameter $\delta\Phi_{\rm x}$ and, simultaneously, applying a suitable sequence of microwave pulses $A\cos(2\pi\nu t+\phi)$ with frequency $\nu\simeq\nu_{01}$. A single microwave pulse results in a rotation of the qubit state vector by an angle $\Omega=2\pi\nu_{\rm R} t\sqrt{1+(\delta/\nu_{\rm R})^2}$ on the Bloch sphere, where $t$ is the pulse duration and the Rabi frequency $\nu_{\rm R} \equiv \nu_{\rm R}(A)$ is a function of the pulse amplitude $A$. The relative phase $\phi$ of the pulse and the detuning $\delta \equiv \nu-\nu_{01}$ determine the rotation axis $\mathbf{v} \equiv (v_1,v_2,v_3) = (\nu_{\rm R}\cos\phi, \nu_{\rm R}\sin\phi, \delta)/ \sqrt{\nu_{\rm R}^2+\delta^2}$. Mathematically, we can describe this rotation with the matrix
\begin{widetext}
\begin{equation}
   \tilde{R}_{\phi,\delta}(\Omega)\equiv\tilde{R}_{\mathbf{v}}(\Omega)=
   \begin{bmatrix}
      \cos\Omega+v_1^2(1-\cos\Omega) & \;\; &
      v_1v_2(1-\cos\Omega)-v_3\sin\Omega & \;\; &
      v_1v_3(1-\cos\Omega)+v_2\sin\Omega \\
      v_2v_1(1-\cos\Omega)+v_3\sin\Omega & &
      \cos\Omega+v_2^2(1-\cos\Omega) & &
      v_2v_3(1-\cos\Omega)-v_1\sin\Omega \\
      v_3v_1(1-\cos\Omega)-v_2\sin\Omega & &
      v_3v_2(1-\cos\Omega)+v_1\sin\Omega & &
      \cos\Omega+v_3^2(1-\cos\Omega)
   \end{bmatrix}
   .
   \label{eqn:rotationmatrix}
\end{equation}
\end{widetext}
For the case $\delta=0$ and $\phi=0$, Eq.~(\ref{eqn:rotationmatrix}) describes a rotation by an angle $2\pi\nu_{\rm R}t$ about the $x$-axis. When introducing a finite relative phase $\phi$ the orientation of the rotation axis within the $x,y$-plane changes. Finite detuning results in a change of the rotation angle and a tilt of the rotation axis out of the $x,y$-plane. In the absence of any microwave radiation ($\nu_{\rm R}=0$) the qubit evolves freely, i.e., its state vector rotates about the $z$-axis of the Bloch sphere with a frequency $\Omega_{\rm free} = 2\pi\delta\cdot t$ (since we work in a frame rotating with frequency $\nu$). The corresponding rotation matrix $\tilde{R}_z(\Omega_{\rm free})$ is obtained from Eq.~(\ref{eqn:rotationmatrix}) by choosing $\mathbf{v}=\mathbf{v}_z\equiv(0,0,1)$.  

%
\begin{figure}[b]
\centering{\includegraphics[width=86mm]{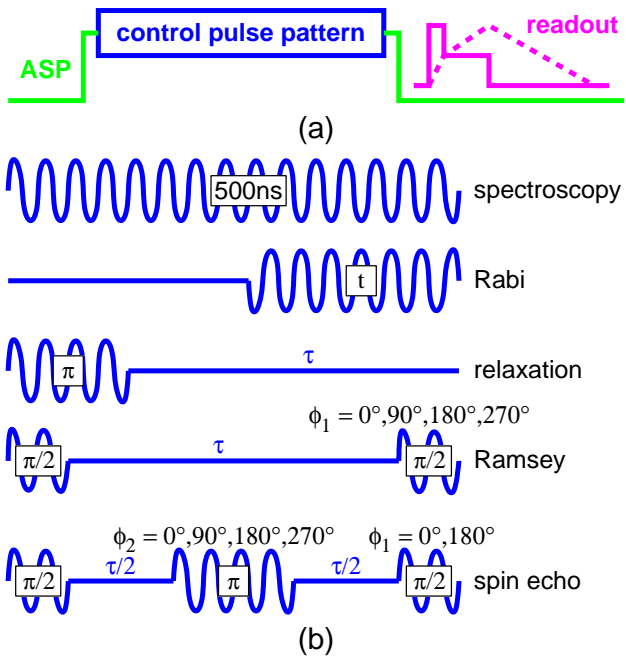}}
\caption{(Color online) (a) Adiabatic shift pulse (ASP) readout. 
         (b) Microwave control pulse patterns for the experiments discussed in
         this work. The boxed values denote either the pulse duration $t$ or
         the corresponding rotation angle of the qubit state vector on the
         Bloch sphere. Free evolution times are denoted by the symbol $\tau$.
         In the multi-pulse sequences, $\phi_1$ and $\phi_2$ are the pulse 
         phases relative to the initial pulse necessary for the phase-cycling
         technique.}
 \label{Deppe_PRB2007_Fig2}
\end{figure}

The microwave pulses are applied to the qubit via an on-chip antenna, which is implemented as a coplanar waveguide transmission line shorted at one end. The qubit is coupled inductively to this antenna. From a FastHenry\cite{FastHenry} simulation the mutual inductance between the antenna and the qubit is determined to be $M_{\rm mw,qb}\simeq73\,$fH. The applied microwave radiation is cooled by means of a $10\,$dB and a $3\,$dB attenuator, which are thermally anchored at a temperature of $4\,$K and $100\,$mK, respectively (cf.\ Fig.~\ref{Deppe_PRB2007_Fig1}). The initial qubit state, which in our measurements is always the ground state, is prepared by waiting for approximately $300\,\mu$s. This time is much longer than the energy relaxation time $T_1$ of the system. For the frequency and time domain experiments, the pulse sequences shown graphically in Fig.~\ref{Deppe_PRB2007_Fig2}(b) are used for the qubit manipulation. For the measurements in the frequency domain (microwave spectroscopy) the qubit is saturated to an equilibrium mixed state by means of a sufficiently long microwave pulse followed by the readout. In the time domain, driven Rabi oscillations are recorded by measuring the qubit state as a function of the duration of a single microwave pulse of fixed amplitude $A$. From the measured Rabi frequency $\nu_{\rm R}$ of these oscillations we determine the durations $t_\pi$ and $t_{\pi/2}$ of the \mbox{$\pi$-} and \mbox{$\pi/2$-pulses}, which rotate the qubit state vector by the corresponding angles. In our experiments we typically choose the microwave pulse amplitude such that $t_\pi=2t_{\pi/2}=1/2\nu_{\rm R}\simeq8\,$ns. Then, the energy relaxation time is determined by exciting the qubit with a \mbox{$\pi$-pulse} and, subsequently, recording the decay of $\langle\sigmaop_z\rangle$ as a function of the waiting time. Finally, the evolution of $\langle\sigmaop_{x,y}\rangle$ is probed by the sequence \mbox{$\pi/2$-pulse}--wait--\mbox{$\pi/2$-pulse}--readout (Ramsey experiment) or the sequence \mbox{$\pi/2$-pulse}--wait--\mbox{$\pi$-pulse}--wait--\mbox{$\pi/2$-pulse}--readout (spin echo experiment). For the latter sequence low-frequency phase fluctuations are canceled because of the refocusing effect of the intermediate \mbox{$\pi$-pulse}. As discussed in more detail below, certain beatings of the spin echo signal can be corrected by recording several traces in which the control pulses have different relative phases $\phi$.  

\subsection{Readout of the qubit state}
\label{subsec:Readoutofthequbitstate}

In order to read out its state, the qubit is surrounded by a slightly larger square-shaped aluminum loop containing two Al/AlO$_x$/Al Josephson junctions, a so-called DC~SQUID (cf.\ Fig.~\ref{Deppe_PRB2007_Fig1}). The DC~SQUID is sensitive to the flux difference generated by the persistent currents flowing in the qubit loop. In our design, the DC~SQUID is coupled to the qubit via a purely geometric mutual inductance $M_{\rm SQ,qb}=6.7\,$pH. Differently from other flux qubit designs \cite{Bertet:2005a,Yoshihara:2006a}, there is no galvanic connection between DC~SQUID and qubit in our sample. This is expected to reduce the effect of asymmetry-related issues as well as the detector back-action on the qubit. In fact, we do not find any measurable bias current dependence of the qubit decay time, as recently recently reported for shared-edge designs\cite{Bertet:2005a,Yoshihara:2006a}. The lines used for biasing and reading out the DC~SQUID detector are heavily filtered against noise in the gigahertz range using a combination of copper powder filters and stainless steel ultra-thin coaxial cables at the mixing chamber temperature level. The details of the filtering of the measurement lines are different for resistive- and capacitive-bias schemes and are explained in detail in Sec.~\ref{sec:Capacitivebias:Theintegratedpulsemethod}.

The detection of the qubit state ($|0\rangle$ or $|1\rangle$) is straightforward far away from the degeneracy point. In this region the energy eigenstates coincide with the persistent current states $\left|{\rm L}\right>$ and $\left|{\rm R}\right>$. The flux generated by these currents is then detected with the DC~SQUID. However, in the vicinity of the degeneracy point the qubit eigenstates $|0\rangle$ and $|1\rangle$ cannot be distinguished in this way because they become nearly equal superpositions of $|L\rangle$ and $|R\rangle$. In other words, the expectation value $\langle I_{\rm p}\sigmaop_z\rangle$ of the persistent current circulating in the qubit loop and thus the flux signal vanishes. To circumvent this problem we employ the adiabatic shift pulse method [cf.\ Fig.~\ref{Deppe_PRB2007_Fig2}(a)]. This method exploits the possibility to perform the readout process sufficiently far away from the degeneracy point by applying a control pulse, which adiabatically shifts the qubit in and out of the region around the degeneracy point. In contrast to the quasi-static flux bias at the readout point, which is generated by a superconducting coil located in the helium bath of our dilution refrigerator, the shift pulse is applied to the qubit through the on-chip microwave antenna (cf.\ Fig.~\ref{Deppe_PRB2007_Fig1}).  The total control sequence (cf. Fig.~\ref{Deppe_PRB2007_Fig2}) for initialization, manipulation, and readout of the qubit can be summarized as follows: First, the qubit is initialized in the ground state at the readout point far away from the degeneracy point by waiting for a sufficiently long time. Then, a rectangular (0.8\,ns rise time) adiabatic shift pulse together with the microwave control sequence is applied to the qubit via the on-chip antenna. In this way the qubit is adiabatically shifted to its operation point, where the desired operation is performed by means of a suitably chosen microwave pulse sequence. Finally, immediately after end of the microwave pulse sequence, the qubit is adiabatically shifted back to the readout point, where the state detection is performed. Note that, in order to avoid qubit state transitions, the rise and fall times of the shift pulse have to be long enough to fulfill the adiabatic condition but also short enough to avoid unwanted relaxation processes.

In the readout process performed right after shifting the qubit back to the readout point, the DC~SQUID is biased with a current pulse of an amplitude just between the two switching currents corresponding to the qubit states.  Depending on the actual qubit state, the DC~SQUID detector either remains in the zero-voltage state or switches to the running-phase state. Only in the latter case a voltage response pulse can be detected from the readout line. The DC~SQUID voltage signal is amplified by means of a room temperature differential amplifier with an input impedance of $1\,\text{M}\Omega$ against a cold ground taken from the mixing chamber temperature level.

The DC~SQUID is shunted with an Al/AlO$_x$/Al on-chip capacitance $C_{\rm sh}=6.3\pm0.5$\,pF. This capacitance also behaves as a filter and, in combination with the other biasing elements (either a resistor or a resistor-capacitor combination), creates the electromagnetic environment of the qubit. Previous studies\cite{vanderWal:2001a} have shown that a purely capacitive shunt results in a much smaller low-frequency noise spectral density compared to an $RC$-type of shunt. This is crucial since one has to reduce the environmental noise as much as possible in order not to deteriorate the qubit coherence times. The resistive part of the biasing circuit helps to damp modes formed by the shunted DC~SQUID and the parasitic inductance/capacitance of its leads. Ideally, it should be placed as close as possible to the shunted DC~SQUID.

Finally, after a typical ``single-shot'' measurement sequence as described above, the response signal of the DC~SQUID is binary (zero or finite voltage state), depending on the qubit state. In the experiments we measure the switching probability $P_{\rm sw}$, which is the average over several thousands of single-shot measurements. For a proper bias current pulse height the qubit state is encoded in the value of the switching probability. In the best case, the ground state would correspond to $P_{\rm sw}^{\rm min}=0\%$ and the excited state to $P_{\rm sw}^{\rm max}=100\%$, or vice versa. In reality, however, due to noise issues the actual visibility $P_{\rm sw}^{\rm max}-P_{\rm sw}^{\rm min}$ is usually significantly smaller than $100\%$.

\section{Capacitive bias: The integrated pulse method}
\label{sec:Capacitivebias:Theintegratedpulsemethod}

In this section, we present our qubit readout scheme, which is based on a capacitive instead of the standard resistive bias for the DC~SQUID. We also refer to this method as the integrated pulse method because the voltage pulse sent to the DC~SQUID detector is the time integral of the desired current bias pulse. Furthermore, the consequences arising for the filtering of the DC lines are discussed. We note that a similar readout method, also based on a capacitive bias, has been proposed in an effort to design and implement a switching detector for Cooper-pair transistor and Quantronium circuits\cite{Walter:2004a,Walter:2007a}. However, there are remarkable differences to the work presented here. Firstly, because of the on-chip shunting capacitor we do not encounter the complication of frequency-dependent damping. Even for low impedances the readout DC~SQUID in our experiments remaines underdamped (DC~SQUID quality factor $Q\simeq4$ at approximately $50\,\Omega$). Secondly, we are able to measure the coherent dynamics of our flux qubit (cf.\ Sec.~\ref{sec:Experimentalresultsanddiscussion}). 

\begin{figure*}
\centering{\includegraphics[width=178mm]{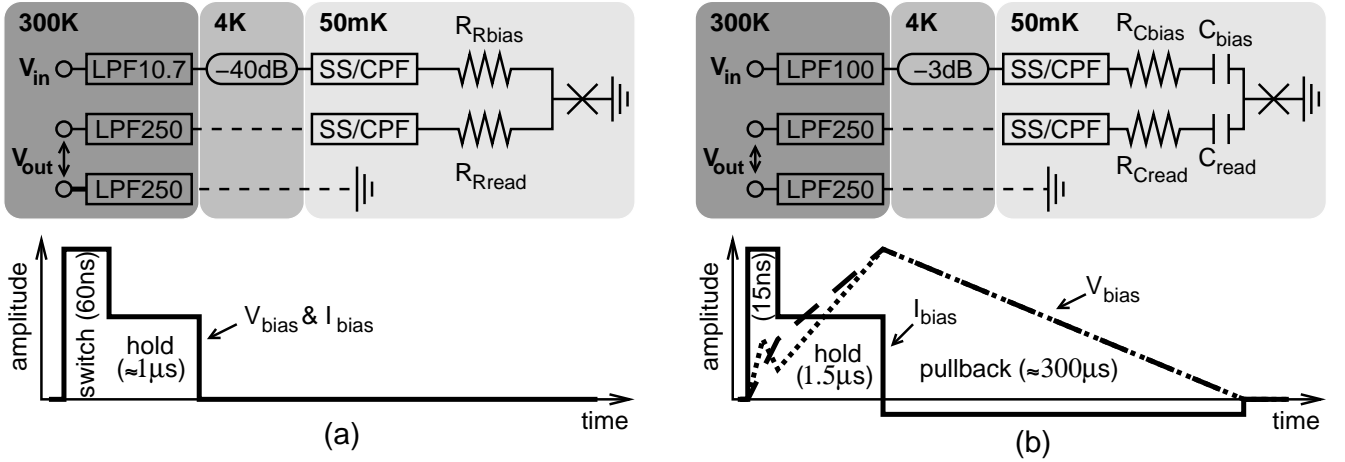}}
\caption{(a) Top: Resistive-bias setup. ``LPF10.7(250)'' denotes a commercial
         low-pass filter with $10.7(250)\,$MHz cutoff frequency. ``SS/CPF''
         represents $1\,$m of ultra-thin ($\varnothing0.33\,$mm) 
         stainless-steel coaxial cable followed by a copper powder filter 
         (bias line: $25\:$cm wire length; readout line: $100\:$cm
         wire length). Qubit, DC~SQUID, and shunting capacitor $C_{\rm sh}$ 
         are indicated with a single cross ($\times$). Solid and broken
         lines represent high-bandwidth semirigid $\varnothing 1.2$\,mm
         CuNi/Nb and narrower bandwidth stainless-steel braided flexible
         coaxial cables, respectively. The bias voltage pulse is attenuated
         by $40\,$dB at $4\,$K. The bias resistors are
         $R_{\rm Rbias}=250\,\Omega+1\,\textrm{k}\Omega$
         and $R_{\rm Rread}=2.25\,\textrm{k}\Omega+3\,\textrm{k}\Omega$
         ($\textrm{on-chip}+\textrm{off-chip}$). 
         Bottom: Switch\&hold readout pulses for the resistive-bias setup.
         Note that $60\:$ns are the width of the portion of switching pulse
         which exceeds the hold level. Only there switching events can
         be induced.
         (b) Top: Capacitive-bias setup: The cutoff frequency of the room
         temperature bias line filter is reduced to 100\,MHz. The voltage
         pulses require a weaker attenuation of $3\,$dB at $4\,$K. The series
         capacitors $C_{\rm bias}=0.5\,$pF and $C_{\rm read}=470\,$pF
         are inserted into the bias and readout lines, forming a
         band-pass filter. Small series resistors ($R_{\rm Cbias}=511\,\Omega$
         and $R_{\rm Cread}=1.5\,\textrm{k}\Omega$) still provide sufficient
         damping of parasitic external modes. Aside from $C_{\rm sh}$,
         all resistors and capacitors are placed off-chip, on the PCB. 
         Bottom: Switch\&hold readout pulses for the integrated-pulse setup:
         The voltage pulse (dashed line) is the time integral of the desired
         current pulse (solid line). An approximately $300\,\mu$s long
         pullback section is required.  In the actual voltage pulses
         (dotted line) another kink is introduced to avoid
         discharging effects of the capacitors.}
 \label{Deppe_PRB2007_Fig3}
\end{figure*}

The readout of the qubit is performed with a switch\&hold measurement technique, which makes use of the hysteretic current-voltage characteristic of the DC~SQUID detector. In the resistive-bias detection scheme, a voltage pulse is generated at room temperature, fed into the DC~SQUID bias line, and then transformed into a current pulse via a cold $1.25\,\textrm{k}\Omega$ bias resistor [cf.\ Fig.~\ref{Deppe_PRB2007_Fig3}(a)]. According to Ohm's law, the voltage pulse must have the same shape as the desired bias current pulse. The latter is composed of a short rectangular switching pulse (of duration $\simeq60\,$ns) immediately followed by a much longer hold pulse ($\simeq1\,\mu$s) of smaller amplitude, as shown in Fig.~\ref{Deppe_PRB2007_Fig3}(a). As already explained in Sec.~\ref{sec:Thesuperconductingquantumcircuit}, the switching pulse height is chosen such that the DC~SQUID either switches to the voltage state or stays in the zero voltage state, depending on the state of the qubit. Hence, the length of the switching pulse determines the time resolution of the switching event detection. The hold pulse level is chosen to be just above the value of the DC~SQUID retrapping current. Consequently, if the DC~SQUID switches into the voltage state during the switching pulse it will not switch back to the zero voltage state during the hold pulse. In this way, the voltage is sustained for a sufficiently long time interval allowing the use of a room temperature amplifier with reduced bandwidth. At the same time quasiparticle generation is minimized.

%

In the resistive-bias experiments the bias lines of the readout DC~SQUID are severely low-pass filtered [cf.\ Fig.~\ref{Deppe_PRB2007_Fig3}(a)]. This is required to reduce the high-frequency noise spectral density $S_\omega(\omega_{01}=E_{01}/\hbar)$, which is responsible for the energy relaxation. In addition, filtering is necessary to eliminate part of the low-frequency noise. In this manuscript, the term low-frequency noise refers to noise which has a large spectral density at frequencies much smaller than the qubit level splitting $\omega_\Delta\simeq4\,$GHz. The dephasing time $T_\varphi$ of the qubit is mainly determined by this low-frequency environmental noise spectral density $S_\omega(\omega\rightarrow 0)$.  Consequently, the cutoff frequency of the low-pass filter should be chosen as low as possible in order to attenuate low-frequency noise strongly. However, a lower limit is set by the requirement that the readout pulse has to be sufficiently short to avoid any deterioration of visibility. For this reason, the filter cutoff cannot be chosen smaller than about $10\,$MHz in the resistive-bias measurements. In other words, in the resistive-bias setup the very low-frequency noise below $10$\,MHz passes unaffected to the sample.

 The integrated pulse readout reduces the effect of low-frequency noise from the DC~SQUID control lines, especially from the bias current line. In a qubit limited by this type of noise, the use of a capacitive bias should improve the dephasing time as compared to the resistive-bias case. The reason is that a band-pass filter instead of a low-pass filter configuration is achieved by replacing the DC~SQUID bias resistor with a $0.5\,$pF capacitor. Consequently, only a narrow frequency band (enough to permit sufficiently fast readout pulses) is allowed to pass, whereas noise at lower and higher frequencies is strongly suppressed. In this way, the rise time of the switching pulses applied to the readout DC~SQUID is reduced by a factor of six ($R$-bias: $60\,$ns, $C$-bias: $10\,$ns). In turn, this should result in a better time resolution and an increased visibility. In actual experiments, we observe an improvement from $20$-$25\%$ for the resistive-bias setup to $28$-$36\%$ for the capacitive-bias configuration (cf.\ inset of Fig.~\ref{Deppe_PRB2007_Fig6}). Since the incoming voltage pulse is differentiated by the bias capacitor, the time integral of the desired current pulse has to be applied as shown in Fig.~\ref{Deppe_PRB2007_Fig3}(b).

\begin{figure}[tb]
\centering{\includegraphics[width=86mm]{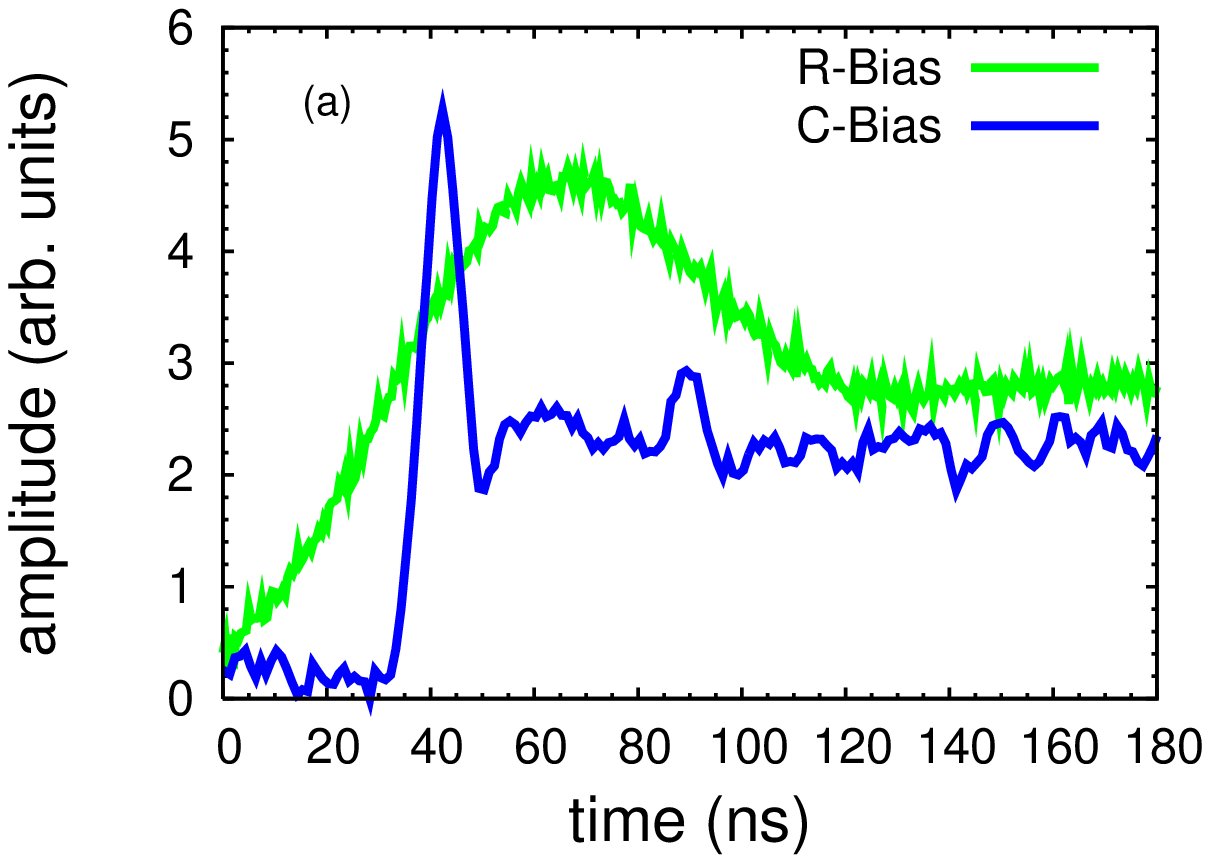}}
 \\
\centering{\includegraphics[width=86mm]{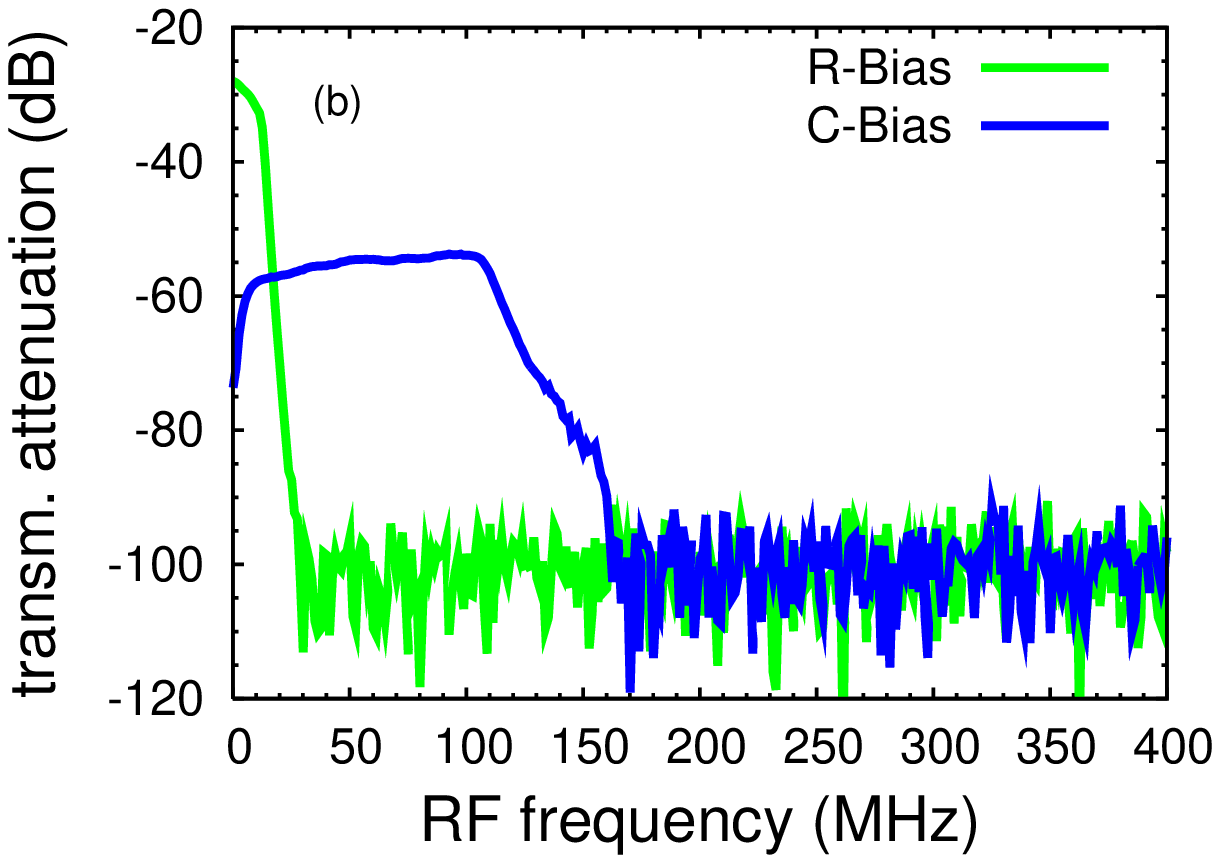}}
\caption{(Color online) Filter characteristics of the DC~SQUID bias line. The green (light gray) and the blue (dark gray) curves are recorded in the resistive and the capacitive bias configuration, respectively. (a) Shape of the bias current pulses. The switching pulse width above the hold level in the capacitive bias setup ($\simeq 15\,$ns) is reduced by a factor of four as compared to the resistive bias setup ($\simeq 60\,$ns). (b) Transmission spectra. In both cases, above $170\,$MHz the transmitted signal is below the noise floor of the network analyzer. In comparison to the resistive-bias setup, stronger attenuation is obtained below $15\,$MHz for the capacitive-bias setup. However, the attenuation is weaker in the $15$-$170\,$MHz window. The width of this frequency range as well as the attenuation factor can be further optimized using a smaller bias capacitor.}
 \label{Deppe_PRB2007_Fig4}
\end{figure}

In Fig.~\ref{Deppe_PRB2007_Fig4}, we compare the pulse shapes and filter characteristics of the two different bias schemes. Evidently, the integrated pulse readout provides a band-pass filtering of the bias line and allows for approximately four times shorter switching pulses than the resistive-bias setup. In part, this is achieved by replacing the $10.7\,$MHz commercial low-pass filter utilized for the resistive-bias readout by a $100\,$MHz filter [cf.\ Fig.~\ref{Deppe_PRB2007_Fig3}(b)]. Note that for the integrated pulse setup the different levels of switching and hold part of the current pulse correspond to different slopes of the voltage pulse in these sections. Since the readout sequence has to be repeated many times, the voltage pulse amplitude has to be reduced back to zero after the end of the hold section. This so-called pullback section has to be much longer ($\simeq 300\,\mu$s in our experiments) than the hold section in order to avoid any switching of the DC~SQUID in the negative current direction. We note that in the experimental implementation the amplitude of the differentiated current pulse falls off the switching to the hold level within a time constant of a few tens of nanoseconds. This is caused by the discharging of the shunting capacitor $C_{\rm sh}$ and can be compensated by introducing a kink in the integrated voltage pulse [cf.\ Fig.~\ref{Deppe_PRB2007_Fig3}(b)]. Moreover, we also place a capacitor $C_{\rm read}$ in the voltage output line. Since small capacitors strongly attenuate the outgoing signal, the value of $C_{\rm read}$ is limited to $470\,$pF.

%

The microwave control signals applied to the sample via the on-chip antenna do not only affect the qubit itself, but also its readout circuitry. This means that the readout signal is usually covered by a wide spectrum of resonances due to parasitic inductances and capacitances of the DC~SQUID leads. In order to avoid parasitic resonances, the DC~SQUID bias resistors used for the resistive-bias measurements are fabricated partially on the sample chip (bias line: $250\,\Omega$ on-chip and $1\,\textrm{k}\Omega$ off-chip; readout line: $2.25\,\textrm{k}\Omega$ on-chip and $3\,\textrm{k}\Omega$ off-chip). In this way, qubit, DC~SQUID, and shunting capacitor are located within a total length scale of about $100\,\mu$m. Thus, most parasitic resonances in the relevant frequency range of a few gigahertz are strongly damped by the bias and readout resistors. The remaining resonances involve the shunting capacitor\cite{Johansson:2006a}, the inductance of the aluminum leads close to the DC~SQUID, possible box resonances, microscopic impurities in the substrate or the junctions, and, of course, the qubit. In the capacitive-bias detection scheme the small bias capacitor pushes parasitic resonances to higher frequencies. In addition, smaller damping resistors are placed in a way that their low-frequency Johnson-Nyquist (thermal) noise is filtered out by the capacitors (bias line: $511\,\Omega$; readout line: $1.5\,\text{k}\Omega$). Because of technical reasons (realization of resistive and capacitive-bias configuration for one and the same qubit), in the capacitive-bias measurements only the shunting capacitance $C_{\rm sh}$ is fabricated on-chip. For the bias and readout resistors/capacitors we use surface mount devices placed on the PCB as close as possible to the sample chip. In this case the on-chip resistors utilized in the resistive-bias measurements are shorted with approximately $1.5\,$mm long gold bonding wires. Fortunately, as it turns out, this does not present a problem for our measurements. Firstly, the spatial enclosing of the sample to a length scale $\lesssim3\,$mm is still sufficient to avoid parasitic resonances. Secondly, although we are limited to a $0.5\,$pF surface mount bias capacitor and although our qubit is energy relaxation limited at the optimal point, the relaxation time is not limited by the high-frequency bias current fluctuations (cf.\ Sec.~\ref{subsec:Energyrelaxation}).

The capacitive setup permits us to obtain a narrow-bandwidth band-pass filter of $150\,$MHz width with a reasonably high center frequency of $90\,$MHz for the DC~SQUID bias line [cf.\ Fig.~\ref{Deppe_PRB2007_Fig4}(b)]. The minimum attenuation is $55\,$dB. This means that we actually only open the frequency band which is absolutely necessary for the readout pulse. In future experiments, the use of smaller on-chip capacitors should allow us to design even narrower band-pass filters with lower minimum attenuation. We finally note that another advantage of the capacitive filter resides in the fact that it is not resistive and, as a consequence, especially suitable for mixing-chamber temperature applications.

\section{Experimental results and discussion}
\label{sec:Experimentalresultsanddiscussion}

In this section, we present the experimental results of spectroscopy (Sec.~\ref{subsec:Switchingdiscriminationandspectroscopy}), energy relaxation (Sec.~\ref{subsec:Energyrelaxation}), and dephasing (Sec.~\ref{subsec:Dephasing}) measurements  performed on our sample. In the discussion we first analyze in detail the results obtained with the capacitive-bias readout. Then, we briefly compare them to those achieved with the conventional resistive-bias method on the same qubit. In the case of the time domain measurements we focus our attention on the region near the degeneracy point of the qubit, where the coherence times are longest, and on the readout point.

However, in the actual experiment the measurements with the well-known resistive-bias scheme are done first. From a technical point of view this has the advantage that the current-voltage characteristics of the readout DC~SQUID can be obtained easily. The reason is that, as described in Sec.~\ref{sec:Capacitivebias:Theintegratedpulsemethod}, only in the resistive-bias configuration very low-frequency signals can be applied to the bias line. 

\subsection{Switching discrimination and spectroscopy}
\label{subsec:Switchingdiscriminationandspectroscopy}

\begin{figure}[tb]
\centering{\includegraphics[width=86mm]{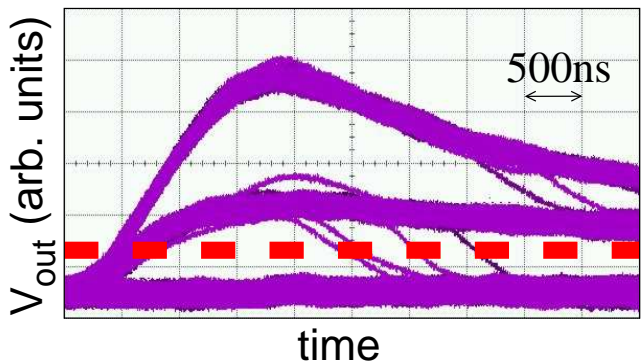}}
\caption{(Color online) Oscilloscope image of the readout DC~SQUID signal
         (solid lines) vs.\ time for the integrated-pulse readout method.
         The switching rate is approximately $50\%$. The broken red line
         marks the detection threshold used to discriminate between 
         zero-voltage (no switching) and finite-voltage (switching)
         events. Note that no retrapping (return to the zero-voltage state)
         occurs before the end of the $1.5\,\mu$s hold pulse. The existence
         of two finite-voltage branches is due to a resonance in the
         current-voltage characteristic of the DC~SQUID. The intermediate
         branch corresponds to the resonance step and the uppermost branch
         to the true superconducting gap.}
 \label{Deppe_PRB2007_Fig5}
\end{figure}

Before starting any experiments on the qubit dynamics, we need to confirm that the considerations made in Sec.~\ref{sec:Capacitivebias:Theintegratedpulsemethod} regarding the feasibility of the capacitive bias-detection scheme are actually valid for the investigated three-junction flux qubit. In other words, we need to verify that the different voltage response signals of the DC~SQUID, which are due to a switching or a non-switching event, can be clearly distinguished from each other. In our sample, this so-called switching discrimination is unambiguous provided that the length of the hold section of the bias pulse is at least $1.5\,\mu$s (cf.\ Fig.~\ref{Deppe_PRB2007_Fig5}). Additionally, we find that we do not detect any spurious switching events on the oscilloscope when the readout pulse does not contain the switching segment. This confirms that the detection of the qubit state occurs only during the short switching pulse.

\begin{figure}[b]
\centering{\includegraphics[width=86mm]{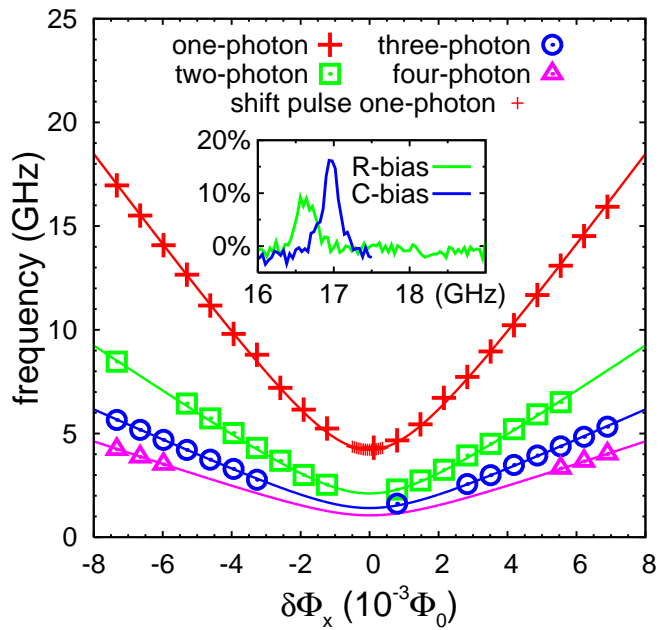}}
\caption{(Color online) Qubit resonance frequency (various symbols)
         obtained from capacitive-bias spectroscopy at $T\simeq50\,$mK 
         plotted vs.\ the external DC flux bias at $T=50\,$mK. Multi-photon
         resonances are observed up to the four-photon process. The solid
         lines are two-level system fits to the measured data. Inset:
         Spectroscopy traces (switching probability vs.\ excitation
         frequency) for a qubit transition frequency $\nu_{01}\simeq17\,$GHz.
         The visibility is two times the maximum amplitude of the resonance
         peak. This is a typical example of the increased visibility of
         the capacitive-bias spectroscopy [blue (dark gray) curve] compared
         to the resistive-bias data [green (light gray) curve].}
 \label{Deppe_PRB2007_Fig6}
\end{figure}

After this initial calibration we measure the qubit energy diagram by means of pulsed microwave spectroscopy (cf.\ Fig.~\ref{Deppe_PRB2007_Fig6}). Each data point in the main body of Fig.~\ref{Deppe_PRB2007_Fig6} corresponds to the center frequency of the observed resonance peak. Let us first look at two typical spectroscopy traces which are displayed in the inset of Fig.~\ref{Deppe_PRB2007_Fig6}. There, the qubit transition  frequency is approximately $17\,$GHz. For the capacitive-bias method the qubit visibility, i.e., two times the height of the spectroscopy peak, is enhanced as compared to the resistive-bias scheme. We find this increased visibility consistently in all spectroscopy and Rabi oscillation data, which has been recorded over a wide range of frequencies (data not shown). More quantitatively, the qubit visibility is $20$-$25\%$ for the resistive-bias method and $28$-$36\%$ for the capacitive-bias scheme. From the main body of Fig.~\ref{Deppe_PRB2007_Fig6} we find a tunnel coupling $\Delta/h=4.22\pm0.01\,$GHz at the degeneracy point. Together with the latter result we can fit the flux-dependence of the qubit resonance signal using a two-level system model. We include one-, two-, three-, and four-photon resonance peak positions into this fit and find a maximum circulating current $I_{\rm p}=360\pm1\,$nA. In order to estimate $\alpha$, we make use of the fact that the two junctions of the readout DC~SQUID have the same layout as the two larger qubit junctions. In this way, we can estimate $I_{\rm c}$ from the measured normal resistance $R_{\rm n}=258\,\Omega$ and gap voltage $V_{\rm g}=390\pm45\,\mu$V of the DC~SQUID. Using the Ambegaokar-Baratoff relation $2I_{\rm c}R_{\rm n}=\pi V_{\rm g}/4$ we obtain $2I_{\rm c} = 1202\pm138$\,nA and, as a consequence\cite{Orlando:1999a},
\begin{equation}
\alpha=\frac{1}{2\sqrt{1-\left(I_{\rm p}/I_{\rm c}\right)^2}}=0.62\pm0.04\:.
\end{equation}
This value is consistent with the result $\alpha_{\rm res}\simeq 0.63$ obtained with the resistive-bias method and deviates only slightly from the design value $\alpha_{\rm design}=0.7$ (cf.\ Sec.~\ref{sec:Thesuperconductingquantumcircuit}). Knowing $\alpha$, we can calculate the Josephson coupling energy $E_{\rm J}/h=299\pm34\,$GHz. Here, the error is dominated by the uncertainty of the gap voltage due to significant fluctuations of the quasiparticle branch of the current-voltage charactristic of the DC~SQUID detector. From previous measurements of junctions made with the same fabrication process\cite{Deppe:2004a}, the charging energy is estimated to be $E_{\rm C}/h\simeq6.4\pm1.6\,$GHz, which gives a ratio $E_{\rm J}/E_{\rm C}=47\pm13$.

\subsection{Energy relaxation}
\label{subsec:Energyrelaxation}

\begin{figure}[tb]
\centering{\includegraphics[width=86mm]{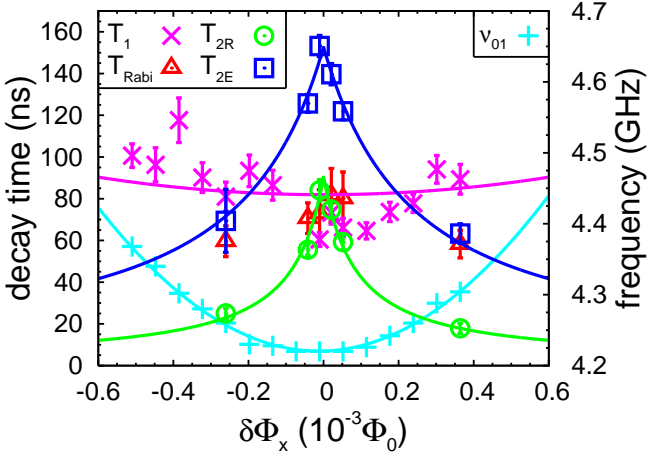}}
\caption{(Color online) Characteristic decay times of the qubit (left scale)
         plotted versus the external flux bias in the vicinity of the qubit
         degeneracy point. The data is obtained from time domain measurements
         with the capacitive-bias method. The magenta (middle gray) crosses
         represent the energy relaxation time $T_1$, the red (dark gray)
         open triangles the Rabi decay time $T_{\rm Rabi}$, the green
         (light gray) open circles the Ramsey decay time $T_{\rm 2R}$,
         and the dark blue (dark gray) open squares the spin echo decay
         time $T_{\rm 2E}$. Also shown are the results of low-power
         spectroscopy (right scale). The light blue (light gray) plus signs
         represent the positions of the resonance peaks. The solid lines
         are fits to the data as discussed in the main text. In addition to
         the data shown in this plot, the results at the qubit readout point
         ($\delta\Phi_{\rm x}=6.007\times 10^{-3}\Phi_0$, 
         $\nu_{01}=E_{01}/h=14.125$\,GHz) are used for the fits.}
 \label{Deppe_PRB2007_Fig7}
\end{figure}

In this section, we evaluate the results of the time domain measurements. We conduct energy relaxation, Rabi, Ramsey, and spin echo experiments using the pulse sequences described in Sec.~\ref{sec:Thesuperconductingquantumcircuit} and Fig.~\ref{Deppe_PRB2007_Fig2}. In our measurements we focus on the flux bias region around the degeneracy point, $\delta\Phi_{\rm x}=\pm6\times10^{-4}\Phi_0$ corresponding to $E_{01} /h \simeq 4$\,GHz, and on the readout point, $\delta\Phi_{\rm x}=-6.007\times 10^{-3}\Phi_0$ corresponding to $E_{01}/h =14.125$\,GHz. All results are summarized in Fig.~\ref{Deppe_PRB2007_Fig7}.

We now discuss in detail the energy relaxation of the flux qubit. According to Bloch-Redfield theory\cite{Bloch:1953a,Redfield:1957a}, which is applicable if the noise is short correlated and weak, the energy relaxation rate is given by\cite{Ithier:2005a}
\begin{eqnarray}
   \Gamma_1 & \equiv & T_1^{-1} = 
   \pi\,S_\omega^{\Gamma_1}(\omega_{01}) = 
   \pi\left(D\frac{\partial\boldsymbol{\omega}}
    {\partial\Phi_{\rm x}}\right)_\perp^2 \;
   S_\Phi^{\Gamma_1}(\omega_{01})
   \nonumber
   \\
   & = & \frac{\pi}{\hbar} \; 
   \left(\frac{\partial\epsilon(\Phi_{\rm x})}{\partial\Phi_{\rm x}}\sin\theta \right)^2\; 
   S_\Phi^{\Gamma_1}(\omega_{01})  \; .
   \label{eqn:gamma1}
\end{eqnarray}
Here, $S_\omega^{\Gamma_1}(\omega_{01})=\frac{1}{2}\left[\widetilde{S}_\omega^{\Gamma_1}(-\omega_{01})+\widetilde{S}_\omega^{\Gamma_1}(\omega_{01})\right]$ is the symmetrized noise spectral density at the qubit transition frequency. The factor $\sin^2\theta=\Delta^2/\big[\Delta^2+\epsilon(\Phi_{\rm x})^2\big]$ can be understood intuitively because the Bloch-Redfield theory is based on a golden rule argument: $T_1$ is certainly related to level transitions of the qubit and $\langle1|\sigmaop_z|0\rangle=\langle0|\sigmaop_z|1\rangle=\sin\theta$ is the transition matrix element due to the interaction between qubit and noise. Also, the importance of the spectral density at the transition frequency becomes clear because only at this frequency state transitions can be induced. In order to calculate $S_\omega (\omega_{01})$ from the flux noise spectral density $S_\Phi (\omega_{01})$, the latter has to be multiplied by the square of the flux-to-frequency transfer function $C_\perp=\big[D(\partial\boldsymbol{\omega}/\partial\Phi_{\rm x})\big]_\perp=\big[\hbar^{-1}\partial\epsilon(\Phi_{\rm x})/\partial\Phi_{\rm x}\big]\sin\theta=(2I_{\rm p}/\hbar)\sin\theta$ because the relaxation rate is determined only by the transverse fluctuations. Close to the degeneracy point we have $\sin\theta \simeq 1$ and the relaxation rate is expected to show a very weak dependence on the externally applied DC flux bias $\delta\Phi_{\rm x}$. Looking at the experimental data of Fig.~\ref{Deppe_PRB2007_Fig7}, we note that although there is some structure in the flux dependence of $T_1$, the relaxation times clearly exhibit a good agreement with the simple Bloch-Redfield model near the degeneracy point. From a numerical fit we derive the flux noise spectral density to be $S_\Phi^{\Gamma_1}(\omega_{01}\simeq\omega_\Delta)=\left[(1.4 \pm 0.1)\times 10^{-10}\Phi_0\right]^2\textrm{Hz}^{-1}$.

\begin{figure}[tb]
\centering{\includegraphics[width=86mm]{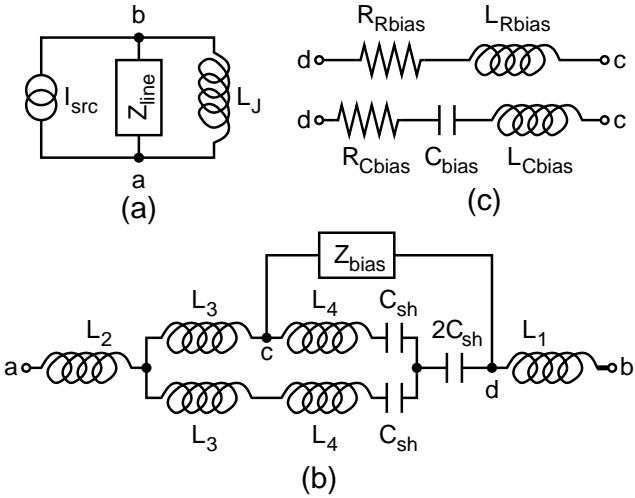}}
\caption{Equivalent circuit of the DC~SQUID bias line. The cables connecting
         the bias resistor to the voltage source are not considered here
         because the corresponding modes are strongly damped. 
         (a) Applying Norton's theorem, the voltage-driven circuit displayed
         in Figs.~\ref{Deppe_PRB2007_Fig1} and \ref{Deppe_PRB2007_Fig3} is
         transformed into an equivalent current-driven circuit. $I_{\rm src}$
         is the effective source current. The DC~SQUID
         is modeled by its Josephson inductance\cite{vanderWal:2001a}
         $L_{\rm J}=2.6\,$nH. $Z_{\rm line}$ is calculated 
         using Kirchhoff's laws. (b) Details of the bias line impedance
         $Z_{\rm line}$. The inductances $L_1=100\,$pH, $L_2=10\,$pH,
         $L_3=60\,$pH, and $L_4=45\,$pH are determined from FastHenry
         \cite{FastHenry} simulations. The bias impedance $Z_{\rm bias}$
         depends on the actual measurement setup. (c) Details of
         $Z_{\rm bias}$ for resistive- (top) and capacitive-bias configuration
         (bottom). $L_{\rm Rbias}=2.7\,$nH and $L_{\rm Cbias}=1.5\,$nH
         are the self-inductances of the on-chip resistor and of the bonding
         wire used to short the on-chip resistor, respectively.}
 \label{Deppe_PRB2007_Fig8}
\end{figure}

\begin{figure}[tb]
\centering{\includegraphics[width=86mm]{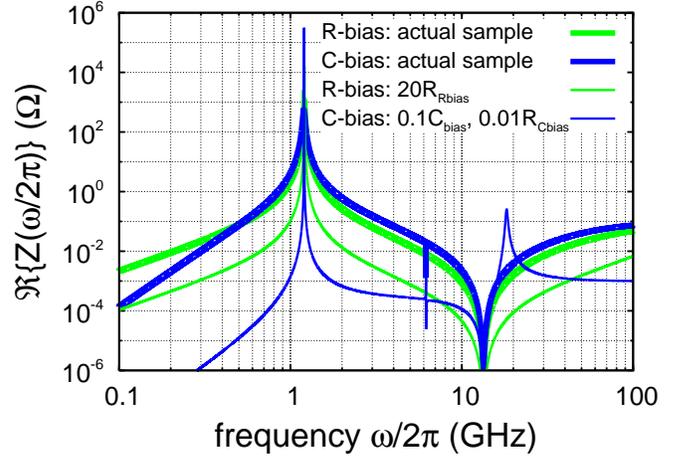}}
\caption{(Color online) Real part of the impedance
         $\Re\left\{Z(\omega/2\pi)\right\}$ of the DC~SQUID bias line
         plotted vs.\ frequency for various configurations. The blue (dark gray) and the green (light gray) lines show the results for capacitive- and resistive-bias configurations, respectively. The actual
         configurations used in the experiments (``actual sample'',
         thick lines) are shown in Fig.~\ref{Deppe_PRB2007_Fig8}. The thin
         lines show suggestions for improved configurations the reduce
         $\Re\left\{Z(\omega/2\pi)\right\}$ in the relevant frequency
         range form $2\,$GHz to $10\,$GHz.}
 \label{Deppe_PRB2007_Fig9}
\end{figure}

With the help of Eq.~(\ref{eqn:gamma1}) we can also estimate the relaxation rate, which would be expected if the impedance $Z(\omega)$ of the DC~SQUID biasing circuitry was the dominating source of energy relaxation. In order to do so, we have to use the voltage noise spectral density $S_V(\omega_{01})$ associated with the real part of the impedance, $\Re\left\{Z(\omega_{01})\right\}$. According to the fluctuation-dissipation theorem, for $\hbar\omega_{01}\gg k_{\rm B}T$, which is true for our experimental situation, we obtain $S_V(\omega_{01}) = 2\hbar\omega_{01} \Re\left\{Z(\omega_{01})\right\}/2\pi$. The flux noise density in the qubit loop due to $S_V(\omega_{01})$ is given by $S_\Phi (\omega_{01}) = M_{\rm SQ,qb}^2 K^2 S_V(\omega_{01})$. Here, $K^2$ is the factor transforming the voltage noise spectral density into a current noise spectral density $S_I(\omega_{01})$ of the current circulating in the readout DC~SQUID. The resulting $S_\Phi (\omega_{01})$ for the qubit is simply obtained by multiplying $S_I(\omega_{01})$ by the quantity $M_{\rm SQ,qb}^2$, which is the square of the mutual inductance between the readout DC~SQUID and the qubit. Hence, we obtain
\begin{eqnarray}
   \Gamma_{1,Z(\omega_{01})} & \equiv & T_{1,Z(\omega_{01})}^{-1}
   \nonumber
   \\
   & = & \frac{\pi}{\hbar}
   \left[\frac{\partial\epsilon(\Phi_{\rm x})}{\partial\Phi_{\rm x}}
   \sin\theta\right]^2 
   \nonumber
   \\
   & & \times M_{\rm SQ,qb}^2 K^2
   \frac{2\hbar\omega_{01} \Re\left\{Z(\omega_{01})\right\}}{2\pi}
   \,,
   \label{eqn:gamma1fromimpedance}
\end{eqnarray}
where\cite{vanderWal:2001a}
\begin{eqnarray}
K^2 & = & \frac{1}{2}\left[\frac{\pi I_{\rm SQ}\tan(\pi f_{\rm SQ})}
          {\omega_{01}\Phi_0}\right]^2 \; \; .
 \label{eqn:gamma1fromimpedance_2}
\end{eqnarray}
In the equation above, $f_{\rm SQ} \equiv \Phi_{\rm SQ}/\Phi_0=0.97 \times 1.5$ is the magnetic frustration of the DC~SQUID due to the flux threading its loop and $I_{\rm SQ}$ is the transport current through the DC~SQUID. From the experimental parameters of the bias circuit (cf.\ Fig.~\ref{Deppe_PRB2007_Fig8}) we first calculate $\Re \left\{ Z(\omega_{01})\right\}$ (cf.\ Fig.~\ref{Deppe_PRB2007_Fig9}). Using Eqs.~(\ref{eqn:gamma1fromimpedance}) and (\ref{eqn:gamma1fromimpedance_2}) we can then estimate the lower limit for the relaxation time caused by the readout process. In this case, we can approximate $I_{\rm SQ}$ with the DC~SQUID switching current $I_{\rm sw}\simeq120\,$nA, obtaining $T_{1,Z(\omega_{01}=\omega_\Delta)}\gtrsim2\,\mu$s  at the degeneracy point. This estimate is already much larger than the actually measured decay time $T_1 = 82\,$ns. Furthermore, in our $T_1$ measurement protocol the bias current is non-zero only at the readout time (cf.\ Fig.~\ref{Deppe_PRB2007_Fig2}). For this reason the readout process itself can only contribute to the limited qubit visibility. The actual DC~SQUID transport current during the $T_1$ measurement sequence is given by asymmetry currents or other noise currents. This means that our experimental situation is characterized by the condition $I_{\rm SQ}\ll I_{\rm sw}$, which implies an even higher expected $T_1$ time. Thus, we conclude that the noise generated by the bias circuitry is not the dominating high-frequency noise source in our experiments. We find a similar result\cite{Kakuyanagi:2007a} for the resistive-bias configuration, where the measured relaxation time $T_1 =140\,$ns is also small compared to the calculated lower bound $T_{1,Z(\omega_{01}=\omega_\Delta)}\gtrsim5\,\mu$s. The slight differences in the $T_1$ values measured with the capacitive- and the resistive-bias readout are mainly due to changes in the microwave setup, which occur during the switching between the different experimental configurations. This process involves the warm-up of the cryostat, sample remounting, opening and closing of microwave connectors, and a new cool-down of the refrigerator.

In conclusion, we find that the current noise from the bias circuitry is not the dominating relaxation source, neither for the resistive- nor for the capacitive-bias detection scheme. Even changes in the high-frequency setup, which are due to the switching between the two methods, affect the $T_1$ times by less than a factor of two. However, in the capacitive setup the measured relaxation rates are more than a factor of $25$ larger than expected from Eq.~(\ref{eqn:gamma1fromimpedance}). The actually dominating noise sources for our qubit are discussed in detail in Sec.~\ref{subsec:Energyrelaxationvsdephasingandnoisesources}. In future works, if these presently dominating noise sources can be reduced to an extent that the bias current noise becomes relevant, the capacitive-bias scheme allows to engineer the circuit impedance and the resulting spectral noise density in an advantageous way. As shown in Fig.~\ref{Deppe_PRB2007_Fig9}, in the relevant frequency band from $2\,$GHz to $10\,$GHz, an impedance (and thus a spectral density), which is almost flat and approximately two to three orders of magnitude smaller than that of the present measurement setup, can be realized by a tenfold reduction of the bias capacitance. In contrast to that, for the resistive-bias readout in the same frequency band such a reduction cannot even be achieved by a twentyfold increase of $R_{\rm Rbias}$. In other words, the bias line noise can be reduced significantly by applying the bias current via a capacitor instead of a resistor. Note that there are no major technical problems hindering a tenfold reduction of the bias capacitor, whereas increasing $R_{\rm Rbias}$ by a factor of 20 can generate significant problems due to heating effects in the bias circuit.

\subsection{Dephasing}
\label{subsec:Dephasing}

In order to determine the dephasing rate $\Gamma_\varphi(\Phi_{\rm x})\equiv T_\varphi^{-1}$ we study the decay of the $\sigmaop_{x,y}$-evolution of the qubit state vector on the Bloch sphere. This requires the use of multi-pulse sequences such as Ramsey and spin echo sequence (cf.\ Sec.~\ref{sec:Thesuperconductingquantumcircuit} and Fig.~\ref{Deppe_PRB2007_Fig2}). The reason for this is that the switching probability of the readout DC~SQUID only reflects the $\sigmaop_z$-component of the Bloch vector. The length of the $\pi/2$- and $\pi$-pulses, i.e., pulses which rotate the Bloch vector about an axis in the $\sigmaop_{x,y}$-plane by the angles $\pi/2$ and $\pi$, respectively, depends on the microwave pulse amplitude and is determined experimentally from driven Rabi oscillations as described in Sec.~\ref{sec:Thesuperconductingquantumcircuit} (data not shown). The azimuthal angle of the rotational axis in the $\sigmaop_{x,y}$-plane is controlled by the relative phase of the pulses. In reality, the microwave control pulses are not perfect: The tipping angle can deviate from the desired value and the frequency can be detuned from the qubit resonance frequency. Neglecting relaxation and dephasing, the Ramsey signal $R_{\phi_1}$ and the spin echo signal $E_{\phi_1,\phi_2}$ can be described by a series of rotations on the Bloch sphere:
\begin{eqnarray}
   E_{\phi_1,\phi_2} &{}\equiv{}& E_{\phi_1,\phi_2}(\delta)
   \nonumber\\[2mm]
   &{}={}&
    \big[ \tilde{R}_{\phi_1,\delta}(\Omega_{\pi/2})
          \tilde{R}_z(2\pi\delta\,t)
    \nonumber\\
   && \times\tilde{R}_{\phi_2,\delta}(\Omega_\pi)\tilde{R}_z(2\pi\delta\,t)
    \tilde{R}_{0,\delta}(\Omega_{\pi/2}) \; \mathbf{x}_0 \big]_z
   \nonumber\\[2mm]
   && \text{ and}
    \label{eqn:phasecycling}
   \\[2mm]
   R_{\phi_1} &{}\equiv{}& R_{\phi_1}(\delta)
   \nonumber\\[2mm]
   &{}={}& \big[ \tilde{R}_{\phi_1, \delta}(\Omega_{\pi/2})
       \tilde{R}_z(2\pi\delta\,t)\tilde{R}_{0,\delta}(\Omega_{\pi/2}) \; 
       \mathbf{x}_0 \big]_z  \; .
   \nonumber
\end{eqnarray}
Here, $\mathbf{x}_0=(0,0,1)$ is the Bloch vector describing the initial state of the qubit. The rotation matrices $\tilde{R}_{\phi,\delta}$ and $\tilde{R}_z$ are those defined in Eq.~(\ref{eqn:rotationmatrix}). They describe the Rabi dynamics and free evolution of the qubit state vector, respectively. $\Omega_{\pi/2}\equiv{\frac{\pi}{2}}\sqrt{1+\left(\delta/\nu_{\rm R}\right)^2}+\alpha^\star$ and $\Omega_\pi\equiv\pi\sqrt{1+\left(\delta/\nu_{\rm R}\right)^2}+\beta^\star$ are the angles by which the $\pi/2$- and $\pi$-pulses determined at resonance actually rotate the qubit Bloch vector in the presence of a detuning $\delta$ and arbitrary pulse length imperfections $\alpha^\star$ and $\beta^\star$. The phase of the initial $\pi/2$-pulse is chosen to be $0^\circ$, that of the final $\pi/2$-pulse $\phi_1$, and that of the intermediate $\pi$-pulse $\phi_2$. In our experiments, we choose the microwave power such that the corresponding Rabi frequency is $\nu_{\rm R}\simeq 65\,$MHz. This results in a $\pi$-pulse length $\tau_\pi = 2\tau_{\pi/2}\simeq 8\,$ns. A more detailed analysis of the measured Rabi oscillations shows that $\alpha^\star,\beta^\star\lesssim 5^\circ$. For perfect pulses ($\alpha^\star=\beta^\star=0^\circ$) with vanishing detuning ($\delta=0$) and zero relative pulse phase ($\phi_1=\phi_2=0^\circ$), the measured signals should be $R_0=-1$ and $E_{0,0}=+1$ independent of $t$. This means that in presence of decoherence we should directly observe the decay envelope, from which the decay time can be readily extracted. However, in practice $\delta$, $\alpha^\star$, or $\beta^\star$ are non-zero, giving rise to time-independent (offsets) and time-dependent (oscillations) changes in the signal. In some cases this makes the proper determination of the decay times rather difficult.

\begin{figure}[tb]
\centering{\includegraphics[width=86mm]{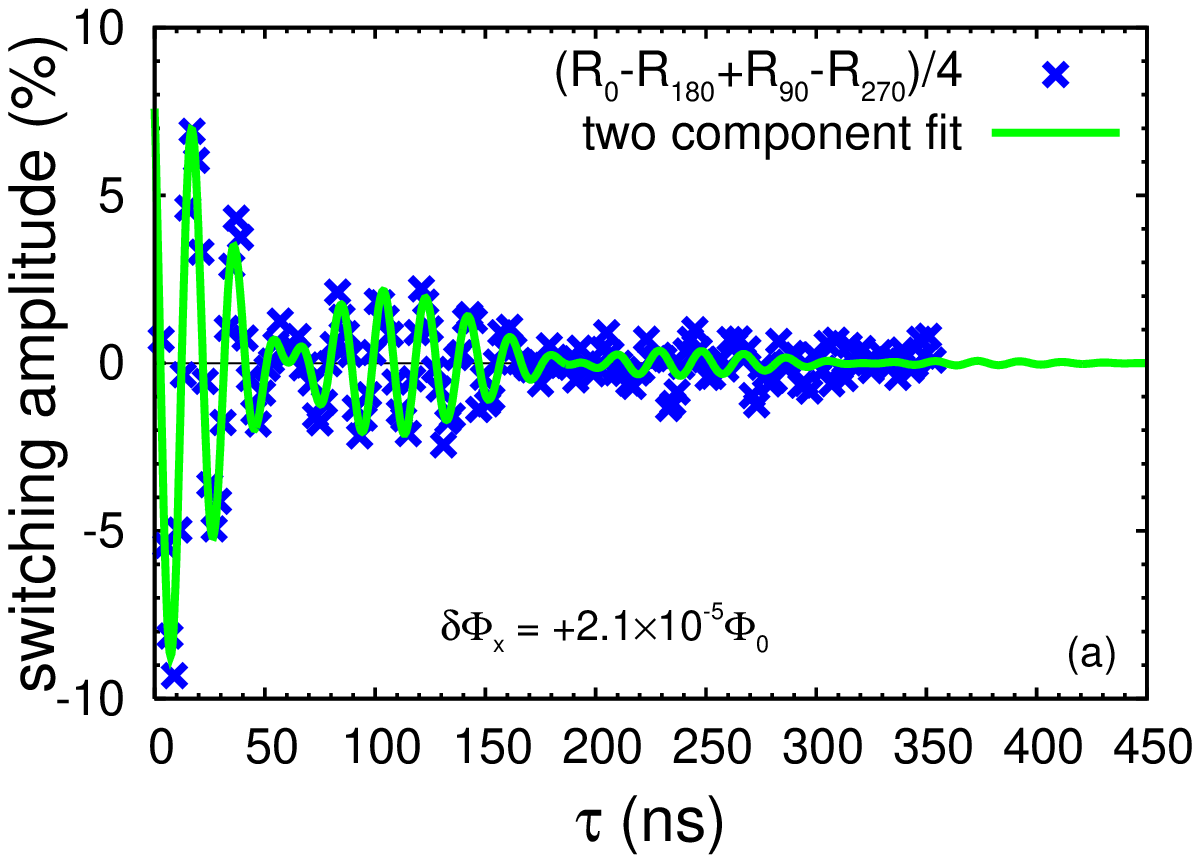}}
\\
\centering{\includegraphics[width=86mm]{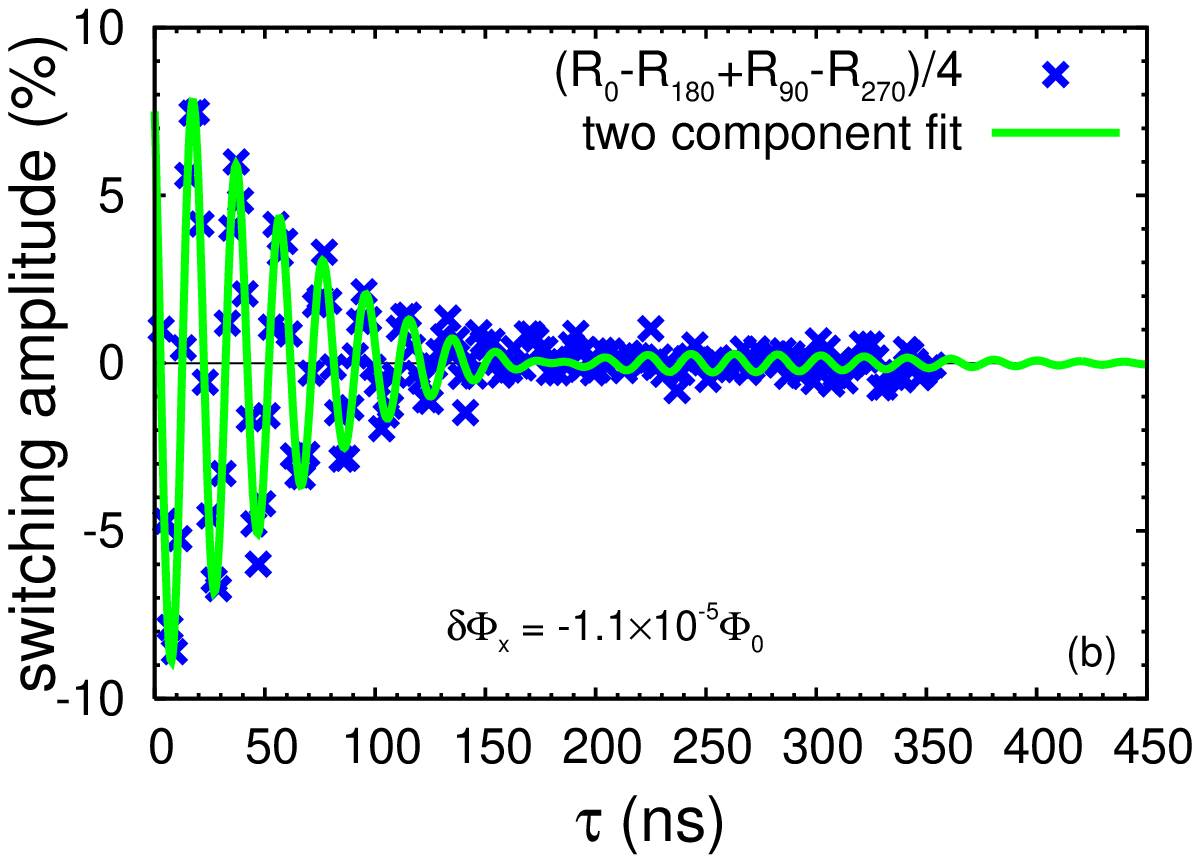}}
\caption{(Color online) Ramsey decay traces measured at two different flux values very close to the degeneracy point using the phase-cycling method. $\tau$ is the free evolution time. For the data [blue (dark grey) crosses] and the traces calculated according to Eqs.~(\ref{eqn:phasecycling}) and (\ref{eqn:ramsey}) the same labeling is utilized. The solid green (light grey) lines are fits to the data using a split-peak model in combination with an exponentially decaying envelope. The observed decay times are (a) $T_{\rm 2R}=75\pm4\,$ns and (b) $T_{\rm 2R}=84\pm5\,$ns.}
 \label{Deppe_PRB2007_Fig10}
\end{figure}

By adding or subtracting traces taken with different relative phases $\phi_1$ and $\phi_2$ the ``ideal pulse'' signals can be restored. This procedure is called phase-cycling method\cite{Kakuyanagi:2007a}. In the case of the Ramsey traces, which are deliberately measured with detuning $\delta_{\rm R}\simeq 50\:$MHz, we can still cancel offsets by considering $R_0-R_{180}$ or $R_{90}-R_{270}$. There are two important offsets: a small one ($\lesssim1\%$) due to pulse imperfections and the calibration offset of the switching probability. The latter is usually chosen to be about $50\%$. In this way, the switching probability is also close to $50\%$, when the qubit is in a $50/50$-mixture of ground and excited state. In order to improve the signal-to-noise ratio we use all four available Ramsey traces
\begin{equation}
R\equiv R(\delta)\equiv R_{0}-R_{180}+R_{90}-R_{270}
   \label{eqn:ramsey}
\end{equation}
for the data analysis (cf.\ Fig.~\ref{Deppe_PRB2007_Fig10}).

\begin{figure}[tb]
\centering{\includegraphics[width=86mm]{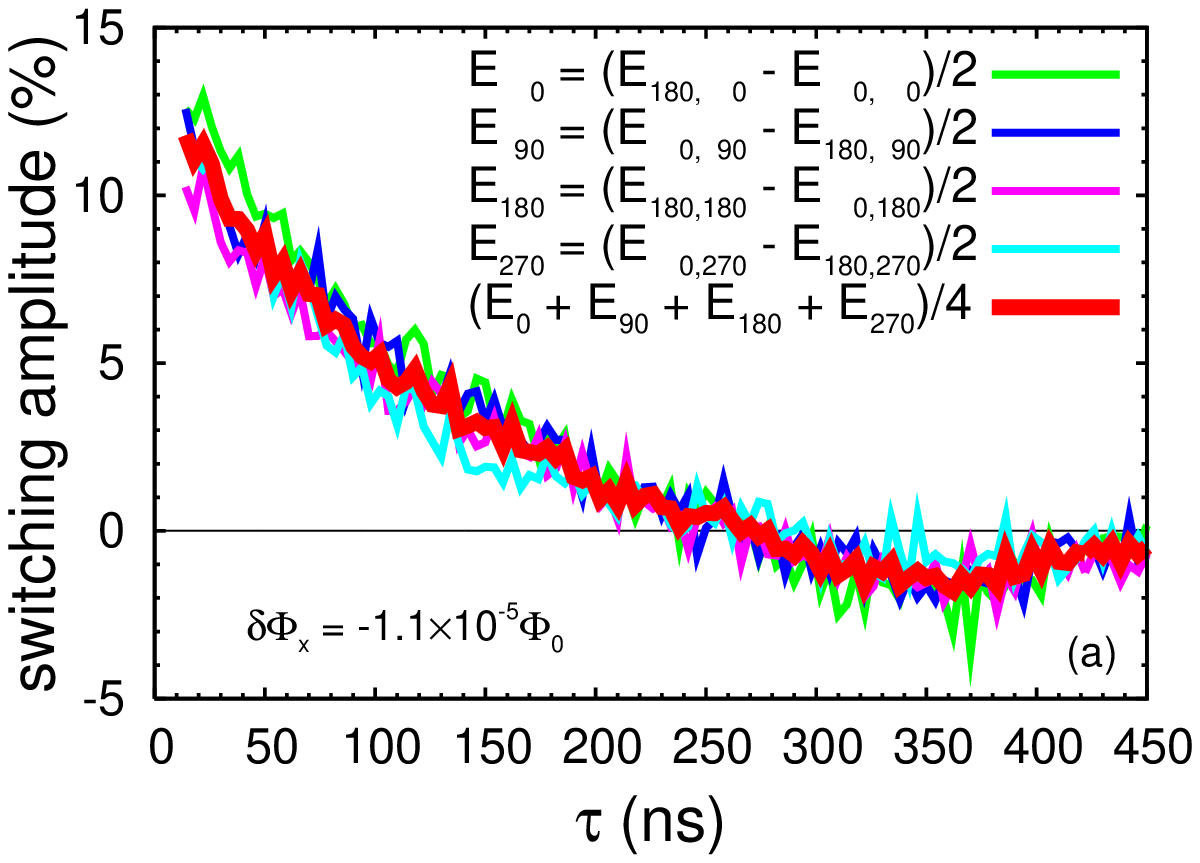}}
\\
\centering{\includegraphics[width=86mm]{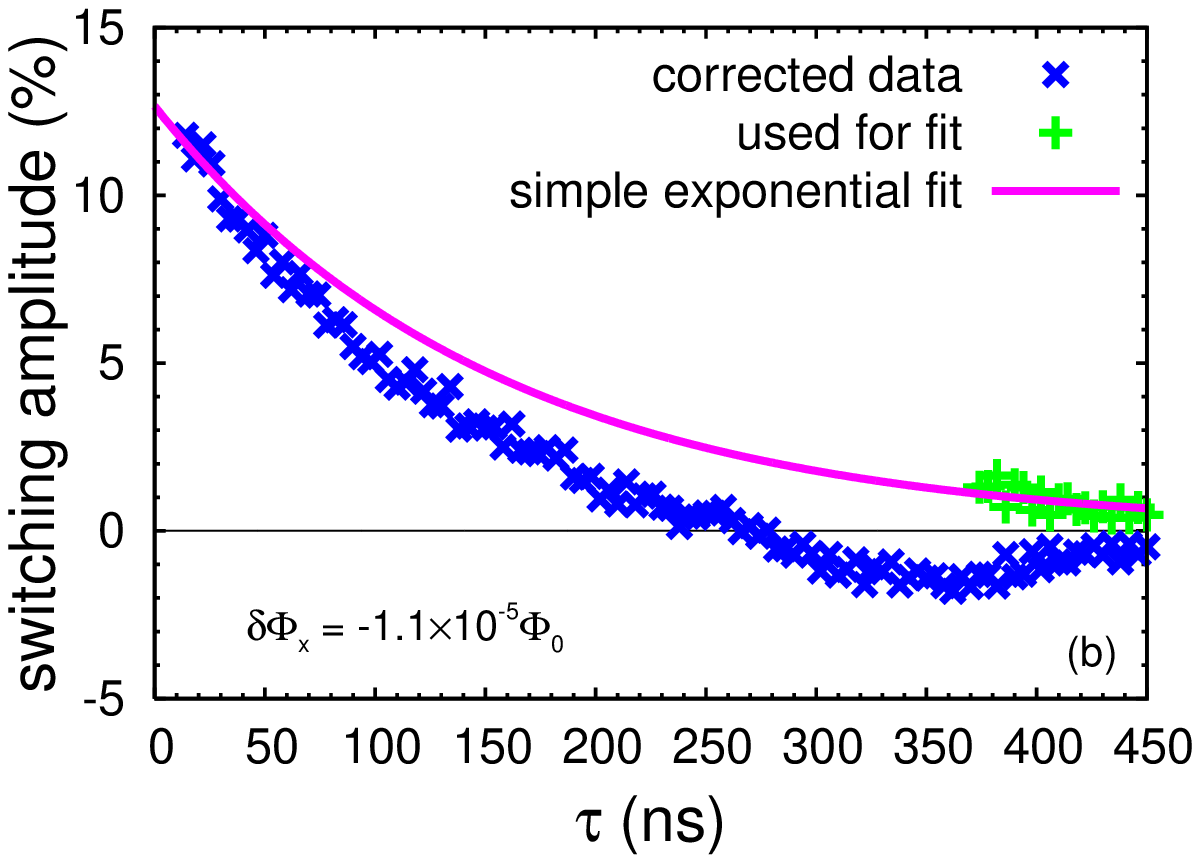}}
\caption{(Color online) Spin echo decay traces measured very close to the degeneracy point using the phase-cycling method. The same labeling is used for the data and for the traces calculated according to Eqs.~(\ref{eqn:phasecycling}) and (\ref{eqn:spinecho}). (a) The aperiodic low-frequency beatings ($\lesssim 2$\,MHz) are independent of the relative phase of the $\pi$-pulse and do not cancel in the corrected trace [thick red (dark gray) line]. Similar to the Ramsey trace $R$, the spin echo difference traces $E_{\phi_1} \equiv {\scriptscriptstyle{}_{(-)}^{\,\,+}}\left(E_{\phi_1,180}-E_{\phi_1,0}\right)$ (thin lines) have a built-in offset correction. (b) Corrected spin echo decay obtained from (a). The solid magenta (medium gray) line is a fit to a simple exponential envelope function. From all data points [blue (dark gray) crosses] only those with absolute values close to the envelope [green (light gray) plus signs] are used for the fit. The amplitude is treated as a fixed parameter which is evaluated from the Ramsey amplitude and the Rabi decay time (Rabi decay data not shown).}
 \label{Deppe_PRB2007_Fig11}
\end{figure}

In the spin echo measurements there is no deliberate detuning ($\delta \lesssim 20\,$MHz) and we are able to keep only the non-oscillating spin echo term using the eight phase combinations
\begin{eqnarray}
   E &{}\equiv{}& E(\delta)
   \nonumber\\
   &{}\equiv{}& E_{0,0}-E_{180,0}-
    E_{0,90}+E_{180,90}
   \nonumber\\
   && {}+E_{0,180} - E_{180,180} 
       - E_{0,270} + E_{180,270}   \,,\quad
   \label{eqn:spinecho}
\end{eqnarray}
which of course decays with some (usually exponential) envelope in the presence of decoherence. With the aid of numerical simulations we confirmed that, for the usual conditions of our spin echo experiments, these corrections are small. In fact, they disappear in the noise floor of the switching probability measurements. However, when looking at the corrected Ramsey and spin echo traces very close to the degeneracy point (cf.\ Fig.~\ref{Deppe_PRB2007_Fig11}), it becomes evident that there are beatings which cannot be corrected by the phase-cycling method. These beatings are discussed in more detail in Sec~\ref{sec:SpinechoandRamseybeatings}.

We now turn to the analysis of the corrected Ramsey and spin echo data, which provides information on the $\sigmaop_{x,y}$- or transverse dynamics of the qubit and, as a consequence, on the spectral density of the phase noise affecting the qubit. The transverse signals have two main decay components. Firstly, there is the energy relaxation already discussed in detail in Sec.~\ref{subsec:Energyrelaxation}, which is obviously still present. Secondly, there are low-frequency random phase fluctuations, which destroy the phase coherence of the qubit. When the energy relaxation is caused by regular high-frequency noise\cite{Makhlin:2003a,Ithier:2005a}, the total rates of the Ramsey ($\Gamma_{\rm 2R}$) and spin echo ($\Gamma_{\rm 2E}$) decay are given by
\begin{equation}
   \Gamma_{\rm 2R}  = \frac{\Gamma_1}{2}+\Gamma_{\rm \varphi R}
   \hspace*{5mm}\mathrm{and}\hspace*{5mm}
   \Gamma_{\rm 2E}  = \frac{\Gamma_1}{2}+\Gamma_{\rm \varphi E}
 \; \; ,
   \label{eqn:spinechorate}
\end{equation}
where the contributions $\Gamma_{\rm \varphi R}$ and $\Gamma_{\rm \varphi E}$ are called the pure dephasing rates. The factor $1/2$ is due to the fact that $\Gamma_1$ is defined as an energy decay, whereas $\Gamma_2$ and $\Gamma_\varphi$ are defined as amplitude decays. Under the assumption of a regular (``white'') noise spectral density $S_\omega(\omega)$ at low frequencies $\omega\simeq0$, we can apply Bloch-Redfield theory again\cite{Bloch:1953a,Redfield:1957a}. In this way, we obtain a simple exponential decay envelope for the pure dephasing and the decay rate is
\begin{eqnarray}
   \Gamma_\varphi^{\rm BR} &=& \pi\,S_\omega^{\rm BR} (\omega=0) =
   \pi\left(D\frac{\partial\boldsymbol{\omega}}{\partial\Phi_{\rm x}}
   \right)^2_\|
   \,S_\Phi^{\rm BR} (\omega=0)
   \nonumber
   \\
   & = & \frac{\pi}{\hbar} \;
   \left[\frac{\partial\epsilon(\Phi_{\rm x})}{\partial\Phi_{\rm x}}\cos\theta
   \right]^2 
   \,S_\Phi^{\rm BR}(\omega=0)\,.
   \label{eqn:blochredfielddephasing}
\end{eqnarray}
In contrast to the energy relaxation, the flux-to-frequency transfer function $C_\| = \big[D(\partial\boldsymbol{\omega}/\partial\Phi_{\rm x})\big]_\|=\big[\hbar^{-1}\partial \epsilon(\Phi_{\rm x})/\partial\Phi_{\rm x}\big]\cos\theta=(2I_{\rm p}/\hbar)\cos\theta$ has to be used since the dephasing rate is determined by the longitudinal fluctuations. The term $-\langle0|\sigmaop_z|0\rangle=\langle1|\sigmaop_z|1\rangle=\cos\theta$ can also be understood intuitively because no level transitions are induced by the low-frequency phase noise. Furthermore, we notice that the flux dependence of the dephasing rate of the qubit is dominated by the factor $\cos^2\theta=\epsilon(\Phi_{\rm x})^2/\big[\Delta^2+\epsilon(\Phi_{\rm x})^2\big]$, which is proportional to $(\delta\Phi_{\rm x})^2$ close to the degeneracy point and approaches unity far away from it. We also notice that the dephasing rate is determined by  $S_\Phi^{\rm BR} (\omega=0)$. This means that $1/f$-noise is of particular importance for the phase coherence of the qubit. However, $1/f$-noise obviously violates the assumption of regularity and one has to calculate the actual noise spectral density from the total accumulated phase\cite{Ithier:2005a}. The result depends on the applied control sequence (Ramsey or spin echo), the order of the coupling (linear or quadratic), and the choice of the frequency cutoff (sharp or crossover to $1/f^2$). It can be Gaussian, exponential of $t^{-\alpha}$ (where $\alpha=3,4$), or even algebraic. Unfortunately, the limited overall data quality of the traces shown in Fig.~\ref{Deppe_PRB2007_Fig11} (short coherence times, switching probability fluctuations, and beatings) does not allow a detailed trace-by-trace analysis of the experimental data. Consequently, we follow a different strategy. We determine the decay rate $\Gamma_2$ from a simple exponential fit of the trace envelope. Because of the aperiodic low-frequency beatings we fit the spin echo traces taking the amplitude as a fixed parameter, which is determined from the Ramsey amplitude and the Rabi decay time. Then, we fit the flux dependence of $\Gamma_2$ with the expression
\begin{eqnarray}
\Gamma_{2}(\delta\Phi_{\rm x}) & = & \frac{\Gamma_1 (\delta\Phi_{\rm x})}{2} + \Gamma_
\varphi^{\rm BR}(\delta\Phi_{\rm x})
  \nonumber  \\
& & + \Gamma_{\varphi}^{1/f}(\delta\Phi_{\rm x}) + \Gamma_{\varphi}^0    \; .
 \label{eqn:gamma2model}
\end{eqnarray}
Here, the pure dephasing rate has a Bloch-Redfield (white flux noise) contribution $\Gamma_\varphi^{\rm BR}$, a Gaussian ($1/f$ flux noise) contribution $\Gamma_{\varphi}^{1/f}$, and a constant (quadratic coupling, flux-independent sources) contribution $\Gamma_{\varphi}^0$. For a $1/f$-noise spectral density
\begin{equation} 
   S_\Phi^{1/f}(\omega)=A/|\omega|
   \label{eqn:1overfspectraldensity}
\end{equation}
one obtains\cite{Ithier:2005a}
\begin{equation}
   \Gamma_{\varphi}^{1/f}(\Phi_{\rm x}) = \frac{1}{\hbar}
   \left|\frac{\partial\epsilon(\Phi_{\rm x})}{\partial\Phi_{\rm x}} 
   \cos\theta\right| \; \sqrt{A\ln2} \; .
   \label{eqn:1overfdephasing}
\end{equation}
Strictly speaking, Eq.~(\ref{eqn:1overfdephasing}) is only true for the spin echo decay. However, the correction necessary for the Ramsey decay rate is only of logarithmic order. Its effect is to increase the impact of the noise by a logarithmic factor, which can be regarded as a constant in our considerations. This means that the $T_{\rm 2R}$-fit produces a Ramsey $1/f$-noise magnitude $A_{\rm R}$ which is by itself unphysical. Nevertheless, $A_{\rm R}$ is useful to understand the filtering effects of the spin echo sequence and estimate the infrared cutoff frequency $\omega_{\rm IR}$ of the $1/f$-noise spectrum.

In our experiments, the Ramsey and spin echo data is taken in the vicinity of the degeneracy point ($\delta\Phi_{\rm x}= \pm 4\times 10^{-4} \Phi_0$, $\nu_{01} = 4.22-4.30\,$GHz) and at the qubit readout point ($\delta\Phi_{\rm x}=-6.007\times 10^{-3}\Phi_0$, $\nu_{01}=14.125\,$GHz). The details of the fitting procedure are different for Ramsey and spin echo data. The Ramsey traces can be fitted properly assuming a simple model in which the main qubit resonance is split into two peaks. Both peaks have equal amplitudes and are detuned by the quantities $\delta_1$ and $\delta_2$ from the main resonance. Using $R(\delta)$ of Eq.~(\ref{eqn:ramsey}), the total decay function consists of beating oscillations $R_{\rm fit}\equiv\big[R(\delta_1)+R(\delta_2)\big]/2$ multiplied by a simple exponential decay envelope. The observed peak splittings $\left|\delta_1-\delta_2\right|$ vary between $3$ and $15\,$MHz around the degeneracy point. At the readout point no splitting is observed. In contrast to the Ramsey beatings, the spin echo beating cannot be fitted using a phenomenological model. From Fig.~\ref{Deppe_PRB2007_Fig11}(a) one can see that only the smaller wiggles are canceled by the phase-cycling method. The long-time-scale beatings coincide for all pulse sequences regardless of the relative phase shifts. For this reason these beatings are not considered as an artifact due to imperfect pulses. Consequently, we fit a simple exponential decay to the envelope of the long-time tail [green (light gray) plus symbols in Fig.~\ref{Deppe_PRB2007_Fig11}(b)] of the total spin echo trace [blue (dark gray) crosses in Fig.~\ref{Deppe_PRB2007_Fig11}(b)]. A substantial part of the error bars of $T_{\rm 2E}$ comes from the fact that it is not always easy to define this tail with high accuracy.

The flux dependence of the decay times $T_{\rm 2E} = \Gamma_{\rm 2E}^{-1}$ and $T_{\rm 2R} = \Gamma_{\rm 2R}^{-1}$ are displayed in Fig.~\ref{Deppe_PRB2007_Fig7}. Obviously, the spin echo decay time is considerably longer than the Ramsey decay time. This indicates the presence of low-frequency noise in the system, which is partially canceled by the echo sequence. Fitting the $T_{\rm 2E}$ data with Eq.~(\ref{eqn:gamma2model}), we obtain a $1/f$-flux-noise magnitude $A=\big[(4.3\pm0.7)\times10^{-6}\Phi_0\big]^2$. Right at the degeneracy point we find $T_{\rm 2E}\simeq 2T_1$ and $T_{\varphi{\rm E}}^0\equiv\left(\Gamma_{\varphi{\rm E}}^0\right)^{-1}\simeq2\,\mu$s, i.e., the total qubit coherence is limited by energy relaxation. These results coincide with those obtained with the resistive-bias readout method\cite{Kakuyanagi:2007a}, although the maximum $T_{\rm 2E}$ is by a factor of two larger in the latter case. This is not surprising, considering the $T_1$ limitation and the fact that already the $T_1$ analysis (cf.\ Sec.~\ref{subsec:Energyrelaxation}) suggests an increase of high-frequency noise due to the switching from the resistive- to the capacitive-readout scheme. Furthermore, we obtain a finite Bloch-Redfield contribution $S_{\rm \Phi,E}^{\rm BR}(\omega=0)=\big[(2.1\pm0.1)\times10^{-10}\Phi_0\big]^2\:\textrm{Hz}^{-1}$ to the spectral flux noise density. In the case of the Ramsey decay, we also find that the flux dependence of the decay time $T_{2\rm R}$ is consistent with the existence of $1/f$-noise ($A_{\rm R}=\big[(2.0\pm0.2)\times10^{-5}\Phi_0 \big]^2$) and $T_{\varphi{\rm R}}^0\equiv\left(\Gamma_{\varphi{\rm R}}^0\right)^{-1}\simeq200\,$ns. However, we  did not find a significant Bloch-Redfield contribution $S_{\Phi,\rm R}^{\rm BR}(\omega=0)$ in the entire flux interval from $\delta\Phi_{\rm x}= -6.007 \times 10^{-3} \Phi_0$ to $+4\times10^{-4}\Phi_0$. In contrast, for the spin echo decay a significant part of the low-frequency $1/f$-noise is canceled ($A_{\rm R}/A\simeq5$) by the refocusing effect of the intermediate $\pi$-pulse. 

\subsection{Energy relaxation vs.\ dephasing and noise sources}
\label{subsec:Energyrelaxationvsdephasingandnoisesources}

\begin{figure}
   \includegraphics[width=86mm]{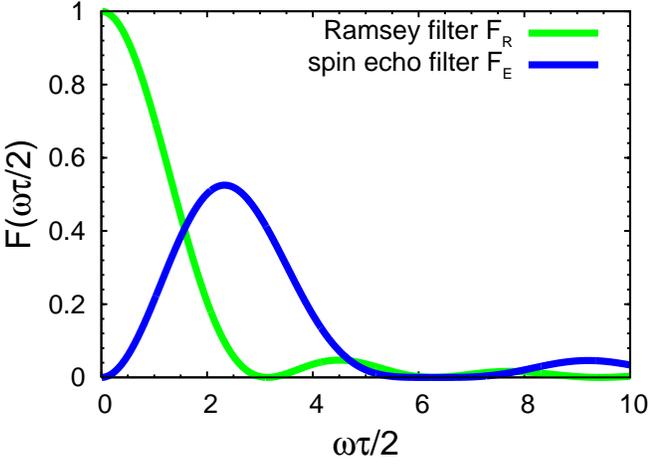}
   \caption{(Color online) Ramsey and spin echo filtering functions.}
   \label{Deppe_PRB2007_Fig12}
\end{figure}

After the quantitative analysis of the influence of the external magnetic flux noise on the qubit decoherence performed in Secs.~\ref{subsec:Energyrelaxation} and \ref{subsec:Dephasing}, we are now ready to put together the results into a general picture in this section. Let us recall that there are three classes of decay times analyzed in this work. The first one consists of the so-called Bloch-Redfield terms ($\Gamma_1$, $\Gamma_\varphi^{\rm BR}$), which can be derived with a golden rule type of argument. The second is due to the impact of $1/f$-noise ($\Gamma_\varphi^{1/f}$) and, finally, there is a contribution $\Gamma_\varphi^0$, which does not depend on the external flux. In order to understand the interplay of these three terms we have to consider their derivation in more detail. Focusing on the dephasing first, under the assumption of a Gaussian distribution of the noise amplitudes we obtain the decay envelope\cite{Ithier:2005a,Martinis:2003a,deSousa:2006a}
\begin{eqnarray}
   f(\tau) &{}={}& \exp\Bigg\{{}-\frac{\tau^2}{2\hbar}
      \left[\frac{\partial\epsilon(\Phi_{\rm x})}{\partial\Phi_{\rm x}}
      \cos\theta\right]^2
   \nonumber\\
   && \hspace{12mm}\times\int\displaylimits_{-\infty}^{+\infty}
    S_\Phi(\omega)F(\omega\tau/2)d\omega\Bigg\}
   \,.
   \label{eqn:dephasingdecayenvelope}
\end{eqnarray}
The filtering functions $F\equiv F_{\rm R,E}(\tilde\omega)$ for the Ramsey and spin echo sequences,
\begin{align}
   F_{\rm R}(\tilde\omega)&=\frac{\sin^2\tilde\omega}{\tilde\omega^2}
   \text{ and}
   \label{eqn:ramseyfilteringfunction}
   \\
   F_{\rm E}(\tilde\omega)&=\frac{\sin^4\left(\tilde\omega/2\right)}
      {\left(\tilde\omega/2\right)^2}
   \label{eqn:echofilteringfunction}
\end{align}
are plotted in Fig.~\ref{Deppe_PRB2007_Fig12}. Here, we use $\tilde\omega\equiv\omega\tau/2$. We can see that the Ramsey sequence is most sensitive to noise close to $\omega=0$. On the contrary, the spin echo sequence is completely insensitive to zero-frequency noise. However, the corresponding filtering function exhibits a maximum at the finite frequency $\omega\approx4/\tau$. Noise at this frequency just cancels the refocusing effect of the intermediate $\pi$-pulse. In other words, $1/f$-noise is strongly reduced by the spin echo sequence, while noise which is white in the angular frequency band between zero and $\omega_{\rm c}\approx12/\tau$ remains unaffected. For typical time scales of our experiments, $\tau\approx T_1,T_2\simeq100\,$ns we find that the critical angular frequency $\omega_c$ is still much smaller than the angular frequency $\omega_\Delta$ corresponding to the qubit gap $\Delta$.

Both the $1/f$-noise decay rate of Eq.~(\ref{eqn:1overfdephasing}) and the Gaussian decay law can be calculated straightforwardly from Eqs.~(\ref{eqn:1overfspectraldensity}), (\ref{eqn:dephasingdecayenvelope}), and (\ref{eqn:echofilteringfunction}). The Bloch-Redfield dephasing rate of Eq.~(\ref{eqn:blochredfielddephasing}) has to be treated carefully: The golden rule argument is only valid for times long compared to $\omega^{-1}$. Since the dephasing is dominated by the low-frequency contribution ($\omega\rightarrow0$) we cannot apply this argument to our time traces (typically $1-450\,$ns long). Instead, we notice that Eq.~(\ref{eqn:blochredfielddephasing}) and a simple exponential decay law can be derived rigorously from Eqs.~(\ref{eqn:dephasingdecayenvelope}) and (\ref{eqn:echofilteringfunction}) assuming a frequency-independent spectral density. Thus, a finite $\Gamma_\varphi^{\rm BR}$ contribution indicates the existence of white noise. If both white and $1/f$-noise are present, near the degeneracy point the peaked shape of $\left(\Gamma_\varphi^{1/f}\right)^{-1}\propto\left(\left|\cos\theta\right|\right)^{-1}$ dominates over the rounded shape of $\left(\Gamma_\varphi^{\rm BR}\right)^{-1}\propto\left(\left|\cos^2\theta\right|\right)^{-1}$. The peaked shape of $T_{\rm 2R}$ and $T_{\rm 2E}$ in Fig.~\ref{Deppe_PRB2007_Fig7} thus represents a clear evidence of the influence of $1/f$-noise on our qubit. Near the readout point $\delta\Phi_{\rm x}=6.007\times10^{-3}\Phi_0$, we have $\left|\cos\theta\right|\simeq\cos^2\theta$ and the influence of the white noise on  $T_{\rm 2E}$ becomes comparable to that of the $1/f$-noise. We find that it is not possible to properly fit the $T_{\rm 2E}$-data (including the readout point), when omitting either the Bloch-Redfield or the $1/f$-term. This means that we find a transition from a $1/f$-noise dominated regime in the direct vicinity of the degeneracy point to a another regime near the readout point, where the influence of $1/f$- and white noise is comparable. In the Ramsey data the $1/f$-noise contribution is clearly dominant because of the singularity at $\omega=0$. For this reason it is difficult to extract the small white noise contribution from the data with a meaningful error bar. The fact that the spin echo sequence strongly reduces $1/f$-noise, but does not affect white noise, allows us to detect the latter.

In order to calculate the true Ramsey decay law due to $1/f$-noise from Eq.~(\ref{eqn:dephasingdecayenvelope}), we need to assume at least an infrared cutoff $\omega_{\rm IR}$. The corresponding decay time is then
\begin{equation}
   \Gamma_{\varphi{\rm R}}^{1/f}(\Phi_{\rm x}) = \frac{1}{\hbar}
   \left|\frac{\partial\epsilon(\Phi_{\rm x})}{\partial\Phi_{\rm x}} 
   \cos\theta\right| \; \sqrt{A\ln\frac{1}{\omega_{\rm IR}\tau}} \; .
   \label{eqn:ramsey1overfdephasing}
\end{equation}
Considering a maximal time trace length of about $\tau_0\simeq300\:$ns, we obtain
\begin{equation}
   \omega_{\rm IR}/2\pi=\left[
      2\pi\tau_0\exp\left(\frac{A_{\rm R}\ln2}{A}\right)\right]^{-1}
      \simeq 0.2\,{\rm Hz}\,.
      \label{eqn:infraredcutoff}
\end{equation}
This result has the same order of magnitude as the inverse time it takes to record a single averaged data point. However, the exponential in Eq.~(\ref{eqn:infraredcutoff}) makes the result exponentially sensitive to measurement errors. Considering that a better time scale for $\left(\omega_{\rm IR}/2\pi\right)^{-1}$ is the time to record an entire decay trace ($\simeq1\,$h), we probably overestimate the infrared cutoff frequency slightly.

Furthermore, we want to point out that the white noise spectral density deduced from the dephasing, $S_\Phi^{\rm BR}$, is of the same order as the Bloch-Redfield spectral density $S_\Phi^{\Gamma_1}(\omega_\Delta)$ inferred from the energy relaxation rates. The energy relaxation rate of Eq.~(\ref{eqn:gamma1}) is calculated with the golden rule assuming weak noise characterized by a smooth spectral density at $\omega_\Delta$. In contrast to the Bloch-Redfield dephasing rate, the long time condition $t\gg\left(\omega_\Delta^{-1}\simeq250\,{\rm ps}\right)$ is fulfilled for our experimental time scales of $1-300\,$ns. The fact that we find $S_\Phi^{\rm BR}\simeq S_\Phi^{\Gamma_1}(\omega_\Delta)\gg S_\Phi^{1/f}(\omega_\Delta)$ suggests that the white noise observed in the $T_{\rm 2E}$-analysis is also limiting the energy relaxation times. Considering that the noise created by the bias circuit cannot explain the observed relaxation rates (cf.\ Sec.~\ref{subsec:Energyrelaxation}) and that $T_1$ is quite sensitive to changes in the microwave setup, the natural candidate for the dominating white noise source is the high-frequency line. In contrast to that, $1/f$-noise can be modeled with an ensemble of microscopic fluctuators close to or within the qubit junctions\cite{Simmonds:2004a}. In Sec.~\ref{sec:SpinechoandRamseybeatings}, we show that we can observe the effect of at least one such fluctuator in our time domain traces.

Let us now compare the performance of our qubit to that of flux qubits with a galvanic coupling to the readout DC~SQUID, which have been measured recently\cite{Bertet:2005a,Yoshihara:2006a}. As a prototypical example, we compare our results to those presented in Ref.~\onlinecite{Yoshihara:2006a}. The most striking similarity is the relaxation limitation at the optimal point. Furthermore, the mutual inductance between the qubit and the high-frequency line as well as the bias line filtering are almost the same. However, the $T_1$-time of our qubit is about one order of magnitude smaller. These observations immediately imply that our qubit is more susceptible to high-frequency noise. We find two main reasons to explain this fact. First, any noise of given amplitude couples more strongly to our qubit because the persistent current is about twice as large as the one in Ref.~\onlinecite{Yoshihara:2006a}. Second, and more important, the total coupling between qubit and DC~SQUID is much lower in our design (no kinetic mutual inductance and smaller DC~SQUID junctions). For manipulations in the vicinity of the optimal point we apply an adiabatic shift pulse (cf.\ Sec~\ref{subsec:Readoutofthequbitstate}) during the qubit operation via the virtually unfiltered high-frequency line. In contrast, in Ref.~\onlinecite{Yoshihara:2006a} the readout pulse, which was sent through the strongly low-pass filtered bias line, causes a flux shift due to the large coupling. This shift does not take place during the qubit operation. In this respect the galvanic coupling method turns out to be advantageous. On the other hand, the controlled application of the shift pulse offers a large flexibility in the experiments. In particular, the possibility to operate the superconducting magnet in the persistent mode is attractive. For this reason, we plan to use an additional low-pass filtered control line for the shift pulse in future experiments.

Finally, other possible noise sources that can deteriorate the qubit relaxation time are dielectric losses in the oxide barrier of the junctions\cite{Simmonds:2004a}. In a three-junction qubit the whole loop acts as a very big fourth junction. Thus, the controlled introduction of a small fourth junction as well as the use of small loop sizes (as it was done in Refs.~\onlinecite{Yoshihara:2006a} and \onlinecite{Bertet:2005a}) might be advantageous because the effective amount of dielectric material in the loop is significantly reduced.

\section{Spin echo and Ramsey beatings}
\label{sec:SpinechoandRamseybeatings}

In this section, we analyze in more detail the beatings which we observe in the Ramsey and spin echo data shown in Figs.~\ref{Deppe_PRB2007_Fig10} and \ref{Deppe_PRB2007_Fig11}. In the previous sections, we already excluded trivial origins of such beatings. Firstly, as pointed out in Sec.~\ref{subsec:Dephasing}, they cannot be caused by imperfect control pulses because we applied the phase-cycling method. Also, because of the spatial enclosing of the sample in combination with the filtering conditions of the measurement lines, parasitic resonances in the experimental setup present only a remote possibility (cf.\ Sec.~\ref{sec:Capacitivebias:Theintegratedpulsemethod}). Finally, as we can see in Fig.~\ref{Deppe_PRB2007_Fig9}, the electromagnetic environment in the immediate vicinity of the qubit does not exhibit any prominent mode with a frequency close to the qubit gap frequency $\Delta/h$. In the following, we suggest a different explanation for the origin of the observed beatings. This explanation exploits the formalism of cavity quantum electrodynamics, which, in general, describes the interaction between atoms and light. In quantum optics, the coupling between an atom and the photonic mode of a cavity can give rise to beating signals\cite{Meunier:2005a,Morigi:2002a}. In our case, the qubit acts as an artificial atom and the cavity is represented by a quantum harmonic oscillator located on the sample chip close to the qubit. As we will see later, the nature of this resonator is microscopic in the sense that its coupling to the microwave drive is much smaller than the qubit-microwave coupling. However, the coupling strength between qubit and resonator is larger than the characteristic decay rates each of them. Under this assumption the system can no longer be described in terms of a spin-boson model. Instead, the qubit-oscillator system has to be treated as a single quantum system. All our arguments given above are supported by numerical simulations, which qualitatively reproduce the main experimental features. Moreover, it is notable that our discussion is not limited to a specific type of qubit or readout method. In addition, when the oscillator is undriven and its thermal population negligible, it qualitatively behaves very similarly to a two-level system. This statement agrees well with our simulations. For completeness, we state that the effect of two-level fluctuators was also addressed in a previous work\cite{Simmonds:2004a}, but with a different focus. 

\begin{figure}[tb]
\centering{\includegraphics[width=86mm]{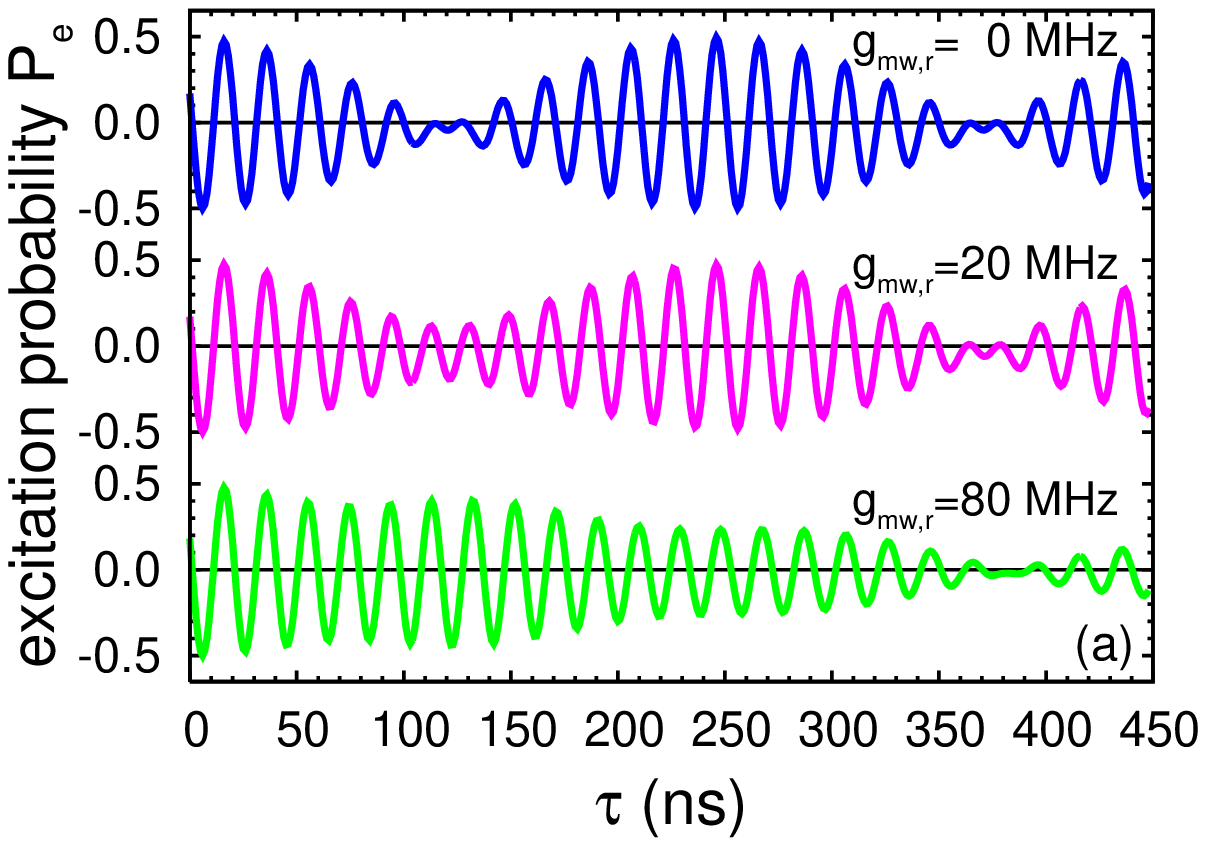}}
\\
\centering{\includegraphics[width=86mm]{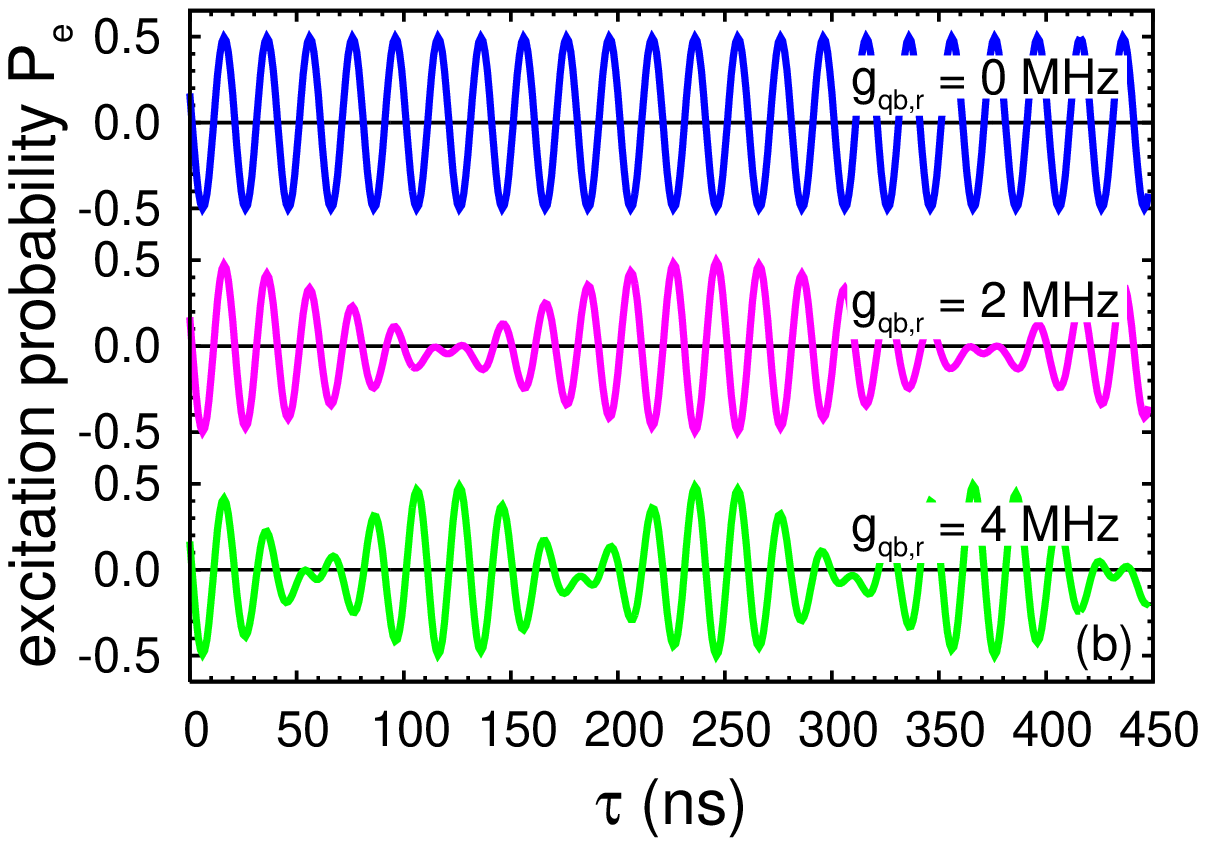}}
\\
\caption{(Color online) Simulated periodic Ramsey beatings due to a qubit-resonator interaction. The resonator (a quantum harmonic oscillator) is truncated to 31 bases. The system parameters are chosen to be similar to those found in the experiment: The qubit persistent current $I_{\rm p}$ is $370\,$nA, the qubit gap frequency $\Delta/h$ is $4\,$GHz, and the resonator frequency $\nu_{\rm r}$ is $4\,$GHz (resonance condition). The maximum free evolution time $\tau$ is $450\,$ns. The microwave excitation frequency is slightly detuned from that of the oscillator ($\nu=\nu_{\rm Ramsey}=4.05\,$GHz). The coupling constants are chosen in a way that the results resemble our experimental data. (a) Variation of the Ramsey signal with increasing $g_{\rm mw,r}$. The  other parameters are $g_{\rm qb,r}=2\,$MHz and $g_{\rm mw,qb}=80\,$MHz for the displayed traces.  (b) Variation of the Ramsey signal with $g_{\rm mw,r}$: The beating frequency increases with $g_{\rm qb,r}$. Our experimental results indicate $g_{\rm qb,r}\simeq2$-$6\,$MHz. The other parameters are $g_{\rm mw,r}=0$ and $g_{\rm mw,qb}=80\,$MHz.}
 \label{Deppe_PRB2007_Fig13}
\end{figure}

\begin{figure}[tb]
\centering{\includegraphics[width=86mm]{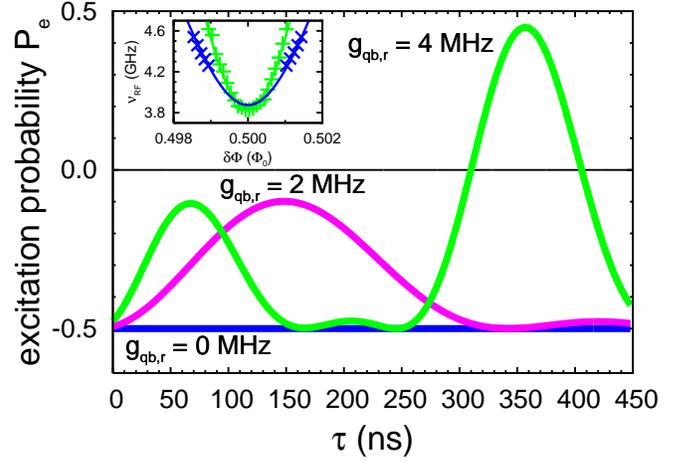}}
\caption{(Color online) Simulated aperiodic spin echo beatings due to the interaction of the qubit with a single harmonic oscillator.  The total free evolution time is $\tau$ and $g_{\rm mw,r}=0$.  Unless otherwise stated, the system parameters are the same as in Fig.~\ref{Deppe_PRB2007_Fig13}.  The microwave driving is resonant with the oscillator ($\nu=\nu_{\rm Echo}=4\,$GHz). Depending on $g_{\rm qb,r}$ the beating spin echo trace crosses or not the zero axis within the chosen time window. If, within the frequency range of interest, the qubit is coupled to several (but not too many) spurious resonators,  their influence has a complex dependence on the external flux bias. For this reason, the presence or absence of the Ramsey and spin echo beatings does not posses any apparent flux bias dependence. However, for a certain flux bias the same beating pattern occurs reproducibly. Such a behavior agrees with our experimental observations (cf.\ Figs.~\ref{Deppe_PRB2007_Fig10} and \ref{Deppe_PRB2007_Fig11}). Inset: Resistive-bias spectroscopy close to the degeneracy point. The green (light gray) plus symbols correspond to the qubit resonance peaks. The blue (dark gray) crosses are consistent with the two-photon process of the blue sideband transition due to the interaction of the qubit with one specific parasitic resonator.}
 \label{Deppe_PRB2007_Fig14}
\end{figure}

The assumption of a harmonic oscillator interacting with the qubit is motivated by spectroscopy experiments performed with the resistive-bias readout. There, we observe a series of peaks which are reminiscent of the two-photon blue sideband transition signal of a qubit-resonator system, in which the harmonic oscillator has a frequency of $\nu_{\rm r} \simeq 4\,\textrm{GHz}\simeq\Delta/h$. These peaks have the same flux-dependence as the normal two-photon peaks, however with a frequency offset (cf.\ inset of Fig.~\ref{Deppe_PRB2007_Fig14}). To model this scenario properly, we make use of the following Hamiltonian, which we write in the qubit eigenbasis:
\begin{eqnarray}
   \frac{\Hop_{\rm qb,r}}{h} &{}={}& 
    \frac{\nu_{01}}{2}\sigmaop_z
    +\nu_{\rm r}\left(\adop\aop+\frac{1}{2}\right)
   \nonumber\\
   && {}+ig_{\rm qb,r}\left(\adop-\aop\right)
    \left(\cos\theta\,\sigmaop_z-\sin\theta\,\sigmaop_x\right)
   \nonumber\\[1mm]
   && {}-g_{\rm mw,qb}\cos\left(2\pi\nu t\right)
   \left(\cos\theta\,\sigmaop_z-\sin\theta\,\sigmaop_x\right)
   \nonumber\\[1mm]
   && {}+ig_{\rm mw,r}\cos\left(2\pi\nu t\right)\left(\adop-\aop\right)
   \:.
   \label{eqn:TLSresontorhamiltonian}
\end{eqnarray}
Here, the first two lines denote an undriven qubit-resonator system, which is similar to the standard Jaynes-Cummings model. The last two lines describe a microwave driving at the frequency $\nu$, which is different for the Ramsey ($\nu=\nu_{\rm Ramsey}$) and spin echo ($\nu=\nu_{\rm Echo}$) experiments. These driving terms are only present within the duration of the control pulses, and they are switched off during the free evolution periods. The operators $\adop$ and $\aop$ are the usual boson creation and annihilation operators for a harmonic oscillator, and $g_{\rm qb,r}$, $g_{\rm mw,qb}$, and $g_{\rm mw,r}$ the coupling constants. The $\left(\cos\theta\,\sigmaop_z-\sin\theta\,\sigmaop_x\right)$-terms appear because of the rotation of the interaction Hamiltonian into the qubit eigenbasis. As a matter of fact, the $\left(\cos\theta\,\sigmaop_z\right)$-term constitutes the main difference to the Jaynes-Cummings approximation and is accurately accounted for in our simulations. Since we are investigating a flux qubit, all interactions are proportional to pairwise products of the qubit current operator $\hat{I}_{\rm qb}$, the oscillator current operator $\hat{I}_{\rm r}$, and the classical microwave driving current $I_{\rm mw}\cos\left(2\pi\nu t\right)$. Also, we can restrict our discussion to the case of pure linear coupling because the microwave power used in our experiments is relatively small.

Based on Eq.~(\ref{eqn:TLSresontorhamiltonian}), we simulate the Ramsey and spin echo sequences at the degeneracy point. We choose similar parameters to those found in the experiments, but neglect the effects of decoherence for simplicity. The results are shown in Figs.~\ref{Deppe_PRB2007_Fig13} and \ref{Deppe_PRB2007_Fig14}. Already at a first glance it is evident that the simulations can reproduce the two main features of the experimental data: There can be periodic Ramsey beatings [Fig.~\ref{Deppe_PRB2007_Fig13}(b)] and, at the same time, there can be quite aperiodic spin echo beatings (Fig~\ref{Deppe_PRB2007_Fig14}). The former are due to an on-resonance qubit-resonator interaction, whereas the latter are caused by the the refocusing $\pi$-pulse, which generates an unwanted evolution in the resonator part of $\Hop_{\rm qb,r}$. When the resonator driving strength $g_{\rm mw,r}$ is increased from zero, the Ramsey beatings are washed out [Fig.~\ref{Deppe_PRB2007_Fig13}(a)]. This indicates that in our experiments the resonator is a point-like microscopic entity (e.g., a magnetic impurity or a dangling bond in one of the qubit junctions), rather than an extended $LC$-circuit\cite{Johansson:2006a}. We use the term ``point-like'' to denote that the resonator must have a small (compared to the qubit) effective area interacting with the external microwave drive. This means that the interaction constant $g_{\rm mw,r}\ll g_{\rm mw,qb},g_{\rm qb,r}$. Consequently, we choose $g_{\rm mw,r}=0$ in the subsequent simulations [cf.\ Figs.~\ref{Deppe_PRB2007_Fig13}(b) and \ref{Deppe_PRB2007_Fig14}]. Nevertheless, the spurious resonator is found to be stable enough to survive at least one thermal cycle. In fact, Ramsey beatings can also be observed in the resistive readout measurements (time domain data not shown, for spectroscopy data cf.\ inset of Fig.~\ref{Deppe_PRB2007_Fig14}). On closer inspection, we find that the resonance condition is very important for our simulation results. This might seem controversial, given that we observe the beatings over a considerable frequency range around the degeneracy point. However, we have to recall that we neglect decoherence here. As a consequence, in the simulations we have much sharper resonance peaks than in the experiments. Furthermore, we only account for a single oscillator. In reality, there are probably several oscillators with slightly different frequencies close to $4\,$GHz and with different coupling constants. It is noteworthy to mention that there is no fundamental restriction for these oscillators to be present only in the frequency range around $4\,$GHz. However, only there the qubit coherence time is long enough to allow for a clear observation of beatings or splittings. The parasitic resonators are also a possible explanation for the fact that the measured relaxation times are much shorter than theoretically expected from the bias current fluctuations (cf.\ Sec.~\ref{subsec:Energyrelaxation}). 

The fact that the simulations suggest a point-like structure as source of the beatings makes it difficult to decide whether we have to assume a harmonic oscillator or a two-level system. Indeed, in the simulations an undriven resonator behaves very similar to a two-level system. Nevertheless, we continue to use the term ``resonator'' in order to keep the language simple and easy to understand. We also want to point out that in the experiments performed with the resistive-bias readout we find similar Ramsey splittings, but no clear spin echo beatings (data not shown). We attribute this to the fact that in the resistive experiments the visibility is worse, while the coherence times are longer. For the rather aperiodic low-frequency spin echo beatings ($\lesssim 2\,$MHz), this results in the tendency to observe the decay envelope only, whereas for the faster, periodic Ramsey beatings ($\simeq 10$\,MHz) the beatings are still visible. Also, the resonance condition discussed above is important for the observability of beatings. The reason is that different oscillators interact with the qubit via different coupling constants. However, beatings cannot be observed unless the resonator frequency falls into the frequency window in which the qubit coherence times are longer than the inverse of the coupling constant. This window is set by the qubit gap, which, due to thermal cycling, shifts by almost $10\%$ from $\Delta_{\rm Rbias}/h=3.9\,$GHz\cite{Kakuyanagi:2007a} in the resistive-bias experiments to $\Delta_{\rm Cbias}/h\equiv\Delta/h=4.22\,$GHz in the capacitive-bias experiments.

\section{Conclusions}
\label{sec:conclusions}

In summary, we have measured the coherent dynamics of a three-Josephson-junction superconducting flux qubit, which is characterized by a gap frequency $\Delta/h=4.22\,$GHz. We used the capacitive-bias detection method, a variant of the switching DC~SQUID readout. This method provides a built-in band-pass filtering of the electromagnetic environment of the qubit, thereby allowing for faster readout pulses and a reduction of low-frequency bias current noise. We use the adiabatic shift-pulse method to probe the qubit close to the degeneracy point and the phase-cycling technique to cancel the effects of control pulse imperfections. Right at the degeneracy point the coherence time of our qubit is limited by energy relaxation and we find $T_1=82\,$ns and $T_\varphi\simeq2\,\mu$s. The flux noise spectral density at the qubit transition frequency is $S_\Phi^{\Gamma_1}(\omega_\Delta)=\big[\left(1.4\pm0.1\right)\times 10^{-10}\Phi_0\big]^2\,{\rm Hz}^{-1}$. In the close vicinity of the degeneracy point, the measured spin echo times exhibit a flux dependence that is expected for the case where flux noise with a $1/f$-frequency spectrum and a magnitude $A=\big[(4.3\pm0.7)\times10^{-6}\Phi_0\big]^2$ is the dominant source of dephasing. Near the readout point, which in our case means far away from the degeneracy point, we also find a significant white noise contribution $S_\Phi^{\rm BR}=\big[\left(2.1\pm0.1\right)\times 10^{-10}\Phi_0\big]^2\,{\rm Hz}^{-1}$. From the comparison between Ramsey and spin echo decay times we estimate an infrared cutoff $\omega_{\rm IR}/2\pi\simeq0.2\:$Hz of the $1/f$-noise spectrum.

On one and the same qubit, we compare the results obtained with the capacitive-bias readout to data taken with the conventional resistive-bias method. We find that the capacitive data shows an increased visibility. On the other hand, a slightly reduced (less than a factor of two) $T_1$-time is observed. The $1/f$-noise magnitude deduced from the spin echo decay is about $\left(10^{-6}\Phi_0\right)^2$ in both setups (cf.\ Ref.~\onlinecite{Kakuyanagi:2007a} for more details on the resistive-bias results). This value was also found in a study where the flux qubit had a galvanic connection to the readout DC~SQUID\cite{Yoshihara:2006a}. Together with the detailed study on the influence of high-frequency environmental bias noise on the energy relaxation time $T_1$, our work clearly shows that external noise generated by the DC~SQUID bias circuitry is not the dominant noise source for either $T_2$- or $T_1$-time. We attribute the difference in the $T_1$-times obtained with the capacitive- and resistive-bias measurements to changes in the high-frequency setup, which have occurred during the experiments in the process of switching between the two methods. Finally, we find that in the frequency range of $2-10\,$GHz, which is most important for flux qubit operation, the capacitive-bias scheme allows one to easily engineer an environment with a low and almost flat noise spectral density of the bias current fluctuations. This is not possible for the resistive-bias readout. In other words, if, in the future, the presently dominating noise sources can be reduced to a level that bias circuit noise becomes the limiting factor, the electromagnetic environment of the capacitive-bias method is advantageous.

Near the qubit degeneracy point, we find beatings both in the Ramsey and the spin echo traces. These beatings constitute an experimental evidence of the interaction of the qubit with one (or a few) harmonic oscillators or two-level fluctuators, which are approximately on resonance with the qubit. We support this argument with the aid of numerical simulations, which qualitatively explain our data. In particular, these simulations indicate that the perturbing oscillators are point-like, i.e., their effective area, and thus their coupling strength to the microwave driving, is small compared to the qubit loop area.

\begin{figure}[tb]
\centering{\includegraphics[width=86mm]{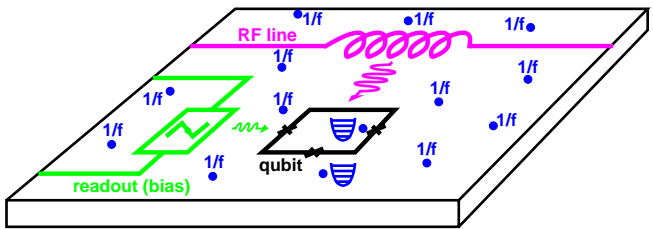}}
\caption{(Color online) Sketch of the relevant noise sources affecting the qubit: We find white noise (arrows), $1/f$-noise [ensemble of blue (dark gray) dots], and a few fluctuators coupled especially strongly to the qubit [blue (dark gray) parabolas]. In our experimental scenario, the high-frequency bias noise of the DC~SQUID [green (light gray) arrow] is much smaller than the noise generated by the high-frequency line [magenta (medium gray) arrow].}
 \label{Deppe_PRB2007_Fig15}
\end{figure}

Finally, comparing the white-noise contribution to the dephasing with the $1/f$-contribution and with the noise responsible for the energy relaxation, we find $S_\Phi^{\rm BR}\simeq S_\Phi^{\Gamma_1}(\omega_\Delta)\gg S_\Phi^{1/f}(\omega_\Delta)$. This means that, according to our experimental results, there are presently two main noise sources limiting the qubit coherence (cf.\ Fig~{\ref{Deppe_PRB2007_Fig15}). The first one is white noise, which can exist up to the qubit gap angular frequency $\omega_\Delta$. This noise represents the main reason for the relaxation limitation at the degeneracy point and also gives a significant contribution to the dephasing of the spin echo signal far away from the degeneracy point. This noise is not generated by the DC~SQUID bias circuit and, as a consequence, most probably couples via the high-frequency line to the qubit. In contrast to that, we observe the effect of $1/f$-noise on the qubit dephasing in the close vicinity of the degeneracy point. Because of the observed beatings in the Ramsey and spin echo signals, we attribute this noise to an ensemble of microscopic fluctuators or resonators located on the sample chip close to or even within the qubit junctions.

\acknowledgments
This work is supported by the Deutsche Forschungsgemeinschaft through Sonderforschungsbereich~631. We wish to thank D.~B.~Haviland, E.~Solano, M.~Thorwart, F.~K.~Wilhelm, and F.~Yoshihara for fruitful and inspiring discussions. We also acknowledge support by the German Excellence Initiative via the Nanosystems Initiative Munich (NIM).

\bibliography{Deppe_PRB2007_v20080919_condmat}

\end{document}